# A 2D Acceptance Diagram Description of Neutron Primary Spectrometer Beams

Leo D. Cussen

Cussen Consulting, 23 Burgundy Drive, Doncaster 3108, Australia

**Email**: leo@cussenconsulting.com

**Telephone**:    +61 401367472

**Abstract**:       Many types of neutron spectrometer use a conventional primary spectrometer consisting of some collimator, a crystal monochromator and a second collimator.  Conventional resolution descriptions use instrument parameter values to deduce the beam character and thence the instrument transmission and resolution.  This article solves the inverse problem of choosing beam elements to deliver some desired beam character at the sample and shows that there are many choices of elements to deliver any given beam character.  Dealing with this multiplicity seems to be a central issue in the search for optimal instrument designs especially if using numerical methods.  The particular approach adopted here is to extend the 2D "Acceptance Diagram" view of the in-scattering plane component of primary spectrometer beams to include horizontally curved monochromators and a variety of collimator types (beamtubes, guides, Soller collimators and radial Soller collimators).  This visual approach clarifies the effect of primary spectrometer variables on the sample position beam and suggests a novel mechanically simple primary spectrometer design offering great flexibility coupled with maximised transmission.

## 1. Introduction

An accurate and complete description of the resolution of neutron scattering instruments is necessary both to analyse the data collected and to design better instruments.  Extensive work has produced mathematical descriptions of the resolution of most significant types of neutron scattering instruments (*e.g.* Small Angle Neutron Scattering diffractometers (SANS), Constant Wavelength Powder Diffractometers (CW PD) and Three Axis Spectrometers (TAS)) but approximations are usually needed to make the description mathematically tractable [*eg* 1-5].  To simplify the mathematics, the beams are usually approximated as having infinite spatial extent and a Gaussian variation of transmission with angle.  This approach works because Gaussians are continuous smooth functions and the product and convolution of Gaussians are also Gaussian.  Even so, the descriptions are so complex that it is now common to resort to Monte Carlo (MC) computer simulations [6-8] of instruments both for instrument design and the derivation of resolution functions.  MC simulations are now also used as the kernel of numerical optimisation approaches to instrument design.  A key issue in such studies is the choice of "quality factor" for comparing results and the most advanced studies recognise the upper limit on performance imposed by Liouville's theorem [*eg* 9].

Many types of neutron scattering instrument (4 Circle Single Crystal Diffractometers (SXD), CW PDs, TAS, *etc*.) use a conventional primary spectrometer (PS) consisting of a source, a collimator, a crystal monochromator and a second collimator. The purpose of this article is to identify the primary spectrometer elements needed to produce a sample position beam with some desired character, a result needed for the analytic optimization of instruments. In some sense this is the inverse of the problem of describing instrument resolution given some choice of beam elements.

A full description of the sample beam requires a distribution of transmission, $\tau$, from the source in a 5D space with the coordinates being horizontal (in-scattering-plane) and vertical spatial ($x$, $y$) and angular ($\gamma$, $\delta$) deviations and wavelength or wave-vector ($\lambda$ or $\kappa$). Such a complete specification makes any mathematical description or discussion complex and visualising or understanding effects in a 5D space is challenging. Conveniently, horizontal and vertical effects are effectively decoupled and can be treated separately because there is effectively no $\gamma$-$\delta$ or $\lambda$-$\delta$ coupling in $\tau$. Vertical divergence effects have been described elsewhere [10-12] and are ignored in this work. Conventional descriptions often remove any spatial variation from consideration by assuming a spatially infinite beam so the description becomes $\tau(\gamma,\kappa)$ or $\tau(\gamma,\lambda)$. The infinite beam approximation is avoided here by considering only the beam at the sample centre ($x_S$=0) and assuming that this represents the beam over the whole sample width accurately enough. MC computer simulations using McStas [6] demonstrate that the description is usually quite accurate. This article develops this 2D $\tau(\gamma,\kappa)$ description of PS transmission to the sample for a variety of collimator types (beamtubes, guides, Soller and Radial Soller) and for horizontally curved or "focussed" monochromators (HFMs). Since one aim here is to describe several types of collimators, the Gaussian profile mathematical and infinite beam approximations are unsuitable. The present work avoids relying upon the Gaussian approximation by using a graphical rather than a strictly mathematical approach. This use of DuMond Diagrams (DDs) [13] or Acceptance Diagrams (ADs) where beam transmission is plotted in an angle : wavelength ($2\theta$-$\lambda$) or angle : wave-vector ($2\theta$-$\kappa$) space has been shown to be valid and to reproduce the known resolution results for CW PDs and TAS [14] but has remained on the fringe of neutron instrument design work. This approach is well matched to the highly developed human capacity for processing visual images so presenting beam descriptions as 2D diagrams should be more immediately accessible and informative than mathematical equations.

The use of vertical monochromator curvature to increase intensity is now widespread and it is becoming common to use HFMs as well. A large number of articles have discussed the resolution effects of HFMs on instruments and presented equations describing their effect [*eg* 15-17]. An earlier 3D ($x$, $\gamma$, $\kappa$) Acceptance Volume (AV) description of in-plane PS transmission [12] dealt with the restricted case of open beamtube collimation with HFMs but the 3D pictures are difficult to interpret and that work incorrectly interpreted guide transmission as acting like a virtual source (although the mathematical results and conclusions appear to be valid anyway).

This document proceeds by defining conventions and symbols and presenting 2D DDs and ADs for a PS using a variety of collimator and monochromator types. It is shown that the usual choice of components leads to over definition of the beam characteristics and mathematical relations are presented in section 6 to describe and exploit this. A novel PS design is described which gives flexibility combined with maximised transmission, simple construction and simple control of beam parameters.

## 2. Problem Outline, schematic, symbols and preliminary results

The conventional PS considered is illustrated schematically in figure 1.

A (Virtual) source, at position "V" has width $\pm W_V$. This may represent the actual source or a slit placed somewhere between the actual source and the monochromator. A crystal monochromator, of width

$\pm W_M$, is centred at position "M", a distance $L_0$ from V, and is oriented at Bragg angle $\theta_M$. The sample, at position "S", is at a distance $L_1$ from M and the line from M to S makes an angle $2\theta_M$ to the line joining V to M. $W_M$ is assumed to be much smaller than $L_0$ and $L_1$ so the monochromator's projected width at source and sample is $\approx \pm W_M \sin\theta_M$. Local right handed Cartesian coordinate sets with $z$ along the beam propagation axis and $y$ vertical are used at the significant points V, M and S in the PS. Individual rays with angular divergences of $\gamma_0$ (between V and M) or $\gamma_1$ (a variation in the scattering angle $2\theta_M$ between M and S) from the beam axes are restricted by beam collimators of half width (HW) $\alpha_0$ or $\alpha_1$ respectively. For collimators of Gaussian transmission profile $\tau(\gamma)$, $\alpha$ represents the full width at half maximum (FWHM). On neutron scattering instruments these angular divergences are invariably small enough that $\sin\gamma \approx \tan\gamma \approx \gamma$.

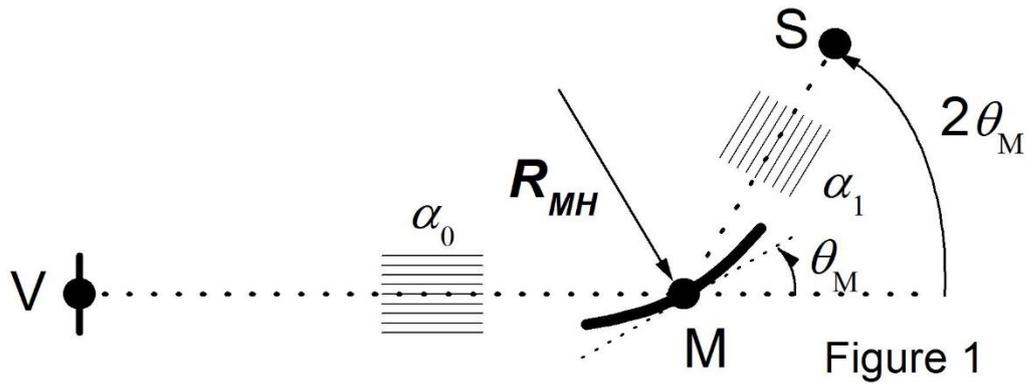

Figure 1

Visually, the AD description is clear and simple, as shown in Figure 2. Expressing the concept in words or mathematics is more complex and turgid. $AD_S$ describes the distribution of transmission (so intensity) in the sample position beam in a 2D wave-vector space with coordinates $\kappa_x$ and $\kappa_z$, so $\tau(\kappa_x, \kappa_z)$ (or equivalently, using coordinates $\kappa[\gamma_1, \Delta\kappa_z/\kappa]$ or the DuMond diagram $\tau[\gamma_1, \Delta\lambda/\lambda]$). $AD_S$ is the product (superposition) of 3 separate ADs associated with individual components - $AD_{\alpha 0}$, $AD_{Mono}$ & $AD_{\alpha 1}$ – and is bounded by an irregular hexagon.

For a mosaic monochromator, $AD_{\alpha 0}$ (for a Soller collimator or guide) is bounded by two parallel lines of slope $-\frac{1}{2}\cot\theta_M$ and crossing the $\kappa_z$ axis at $\pm\frac{1}{2}\alpha_0\cot\theta_M$. The transmission is modulated by the $\alpha_0$ profile. For a radial Soller collimator or open beamtube the slope changes by a factor $(1-L_1/L_0)$.

$AD_{Mono}$ (for a flat mosaic monochromator) is bounded by two parallel lines of slope $-\cot\theta_M$ which cross the $\kappa_z$ axis at $\pm\eta_M\cot\theta_M$. For a curved monochromator the slope changes by a factor $(1-L_1/R_{MH}\sin\theta_M)$. The transmission is modulated by the $\eta_M$ profile.

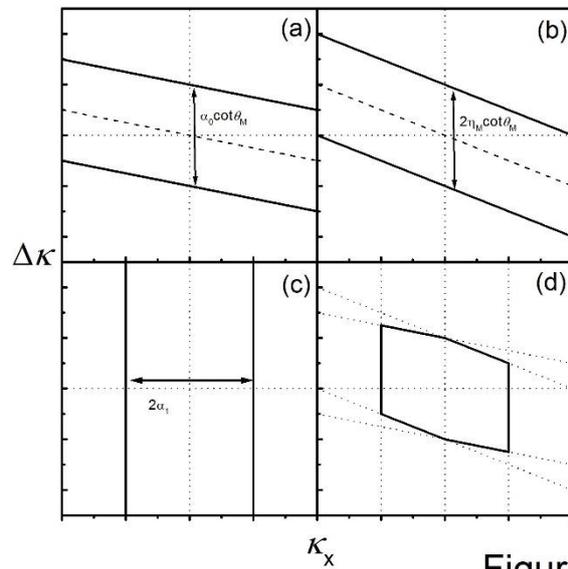

Figure 2

$AD_{\alpha 1}$ just limits the angular width ($\kappa_x \approx \kappa\gamma_1$) at the sample with no effect on the $\kappa$ variation so it is bounded by two lines parallel to the $\Delta\kappa$ axis which cross the $\gamma_1$ axis at $\pm \alpha_1$ with transmission modulated by the $\alpha_1$ transmission profile.

The next sections of this article develop the mathematics needed to demonstrate that the pictures truly represent the situation, to allow the pictures to be drawn accurately and to enable them to be applied to instrument optimisation.

## 2.1 The transmission of individual PS beam elements

The transmission of different types of collimators (Soller collimators, idealised Straight Guides, Radial Soller Collimators and open beamtubes) is conveniently visualised using position–angle Phase Space Diagrams, $\tau(x,\gamma)$ as shown in figure 3. All such collimators are insensitive to small variations in neutron wavelength. Strictly, a guide's transmission angular width is proportional to wavelength but since the wavelength spreads considered here are typically less than 1%, this is a negligible effect. The transverse component of the wave-vector, $\kappa_x$, or neutron momentum is closely proportional to the angular divergence, $\gamma$, for small divergences.

Ideal guides and Soller collimators produce beams where the angular divergence distribution is independent of position, $x$. Radial Soller collimators and open beamtubes produce beams with correlations between position and angular distribution.

- An ideal long guide tube allows full transmission ($\tau=100\%$) up to some finite limit in space (guide width) and in angle ($m\theta_C\lambda$ where $m\theta_C$ characterises the critical angle for the guide mirror coatings) as illustrated in figure 3a. This effect can also be achieved in a short length using a reflecting Soller collimator. The mirror reflections may disrupt any existing angle-spatial correlations in the beam.

- An ideal Soller collimator allows transmission which is spatially uniform with a triangular variation in angle, $\tau(\gamma)$, as shown in figure 3b. Note that this is an idealised representation as would be seen on averaging the beam over a spatial width much larger than a single collimator channel width.

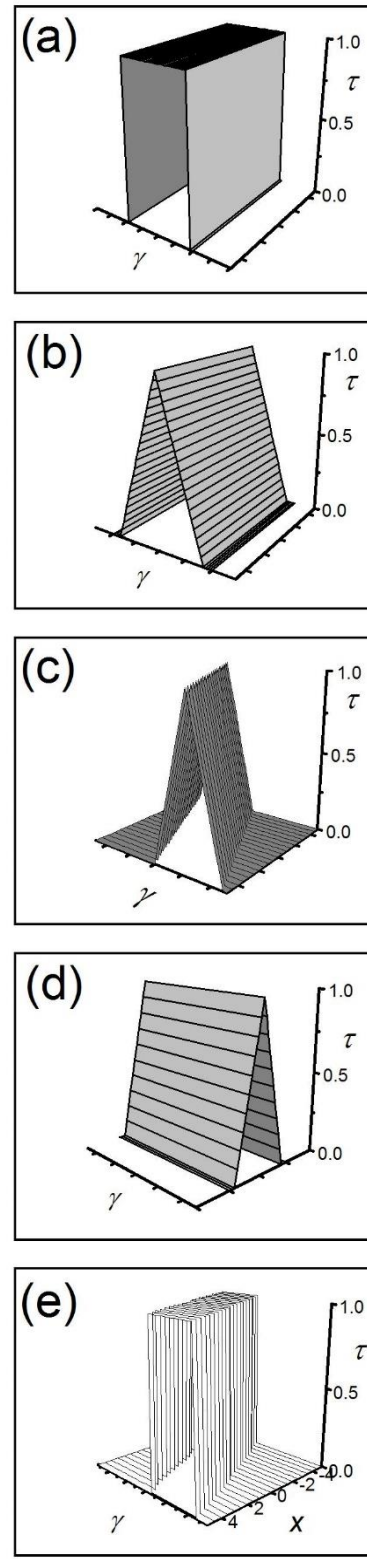

Figure 3

The behaviour of radial Soller collimators depends on whether they are converging or diverging.

- A diverging radial Soller collimator (as would be used between V and M) gives a triangular $\tau(\gamma)$ where the HW, $\alpha$, is determined by the blade separation and length and the centre point of the distribution depends on transverse position $x$ at distance $L_0$ as $\gamma_{Centre} \approx x/L_0$. This is illustrated in figure 3c. Such a collimator effectively limits the source width visible.

- A converging radial Soller collimator (as would be used between M and S) gives transmission independent of $\gamma$ but with a triangular variation with position, $\tau(x)$, as illustrated in figure 3d. Such a collimator effectively limits the beam spatial width at the sample.

For beamtubes, the local angular distribution is rectangular with constant width but the centre of the angular distribution varies systematically with transverse position.

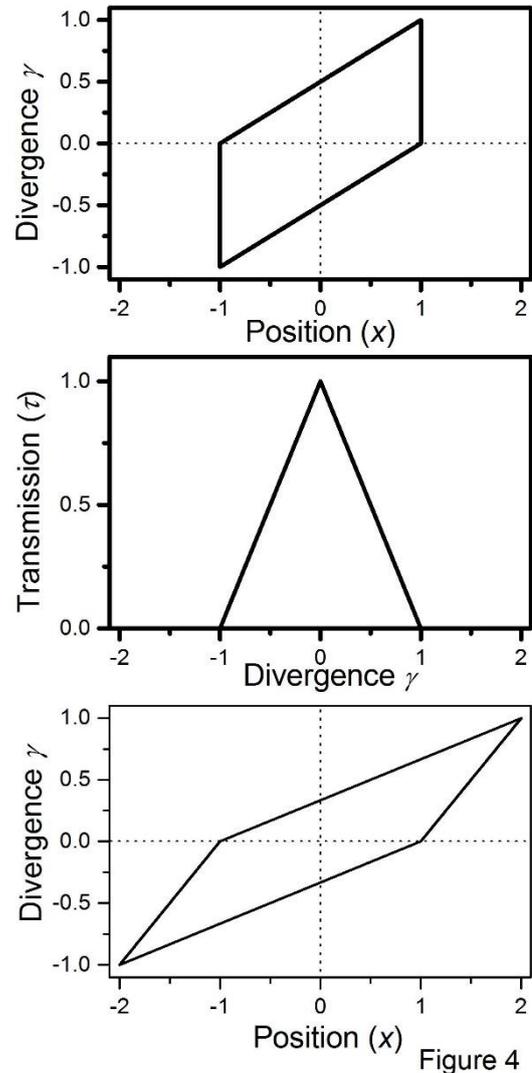

- A diverging open beamtube (or a separated slit pair), as for $\alpha_0$ between V and M, allows uniform full transmission at the monochromator for angles $\gamma_0$ in the range $x_M \sin\theta_M / L_0 \pm W_V/L_0$ as shown in figure 3e. Here $x_M$ is the $x$ position along the monochromator surface. The illustration here is for a slit pair where the first slit is narrow and the second quite wide. For the case where the slits are of equal width, figure 4a shows the $\tau(x,\gamma)$ plot immediately after the second slit, figure 4b shows $\tau(\gamma)$ immediately after the 2$^{nd}$ slit and figure 4c shows the figure 4a phase space diagram sheared as would be seen some distance behind the second slit.

- A converging open beamtube, as for $\alpha_1$ between M and S, gives $\tau(\gamma) = 100\%$ at the sample for angles $(x_S - W_M \sin\theta_M)/L_1 < \gamma_1 < (x_S + W_M \sin\theta_M)/L_1$ where $x_S$ is the $x$ position at the sample. This work only explicitly considers the beam at $x_S = 0$ - it is simple enough to extend the results to finite $x_S$.

Figure 4

## 2.2 Bragg scattering at the monochromator

A crystal monochromator scatters neutrons according to Bragg's Law,

$$n\lambda = 2d_M \sin\theta_M \qquad \text{or} \qquad \kappa = \pi/nd_M\sin\theta_M \qquad \text{(Eq. 1)}$$

This gives the nominal scattered wavelength, $\lambda$, or wave-vector, $\kappa$, where $d_M$ is the monochromator lattice plane spacing and n is an integer (assumed to be 1). Combined with simple geometry, Bragg's Law permits the calculation of the PS transmission as a function of angular and wave-vector divergence. Beams incident on a monochromator usually have some angular spread. Monochromator crystals are usually imperfect with a local "mosaic" or variation in crystallite orientation, $\eta$, often modelled by a Gaussian distribution of FWHM $\eta_M$. Monochromator mosaic permits a variation in scattering angle and in wavelength or wave-vector. A $d$- spacing gradient, $\mu$, a fractional variation in $d_M$ of up to $\pm\mu_M$, may also occur. A $d_M$ gradient allows scattering of different $\lambda$ at a single Bragg angle.

$$\frac{\Delta\lambda}{\lambda} \approx \Delta\theta_M cot\theta_M + \mu \qquad \frac{\Delta\kappa}{\kappa} \approx -\Delta\theta_M cot\theta_M - \mu \qquad \text{(Eq. 2)}$$

The variation in Bragg angle is given by

$$\Delta\theta_M = \gamma_1 - \xi \qquad \text{(Eq. 3a)}$$

$$\Delta\theta_M = \xi - \gamma_0 \qquad \text{(Eq. 3b)}$$

$$\Delta\theta_M = \frac{1}{2}(\gamma_1 - \gamma_0) \qquad \text{(Eq. 3c)}$$

where $\xi$ is the local monochromator crystallite misorientation due to the combined effect of curvature and mosaic. Monochromator curvature can be regarded as "organised" mosaic (or mosaic as chaotic curvature) giving a systematic variation of $\xi$ with position $x_M$ along the monochromator surface.

$$\xi = \frac{x_{M0}}{R_{MH}} + \eta \approx \frac{\gamma_1 L_1}{R_{MH}sin\theta_M} + \eta \qquad \text{(Eq. 4)}$$

where $R_{MH}$ is the monochromator radius of curvature. $x_M$ changes sign in passing the crystal due to the reflection and $x_{M0}$ indicates that the $x$ value before reflection is used. The scattering crystal element must have $\xi = ½(\gamma_1+\gamma_0)$ whence $\gamma_0 = 2\xi - \gamma_1$. To solve the ray tracing equations and proceed with the analysis some restriction must be applied and usually this is to assume a uniform beam of infinite width. Since this assumption cannot be justified for HFMs or converging or diverging collimators or beamtubes, it is replaced here by considering only rays reaching the sample centre, $x_S$=0, when

$$\gamma_1 \approx \frac{x_{M0}sin\theta_M}{L_1} \qquad \text{(Eq. 5)}$$

Often, the beam at $x_S$=0 is representative of the beam over the whole sample width but, if this is likely to be important to the measurement considered, this can be checked either by an MC simulation or by calculating the beam character for $x_S \neq 0$.

### 3. DuMond Diagrams for conventional primary spectrometers

This section presents 2D DuMond diagrams in a $2\theta$-$\lambda$ space illustrating the effect on the sample position beam of the elements of conventional Primary Spectrometers using flat monochromators (figure 5). The reciprocal space formalism of Von Laue is more powerful and general but it is usual to first learn diffraction theory in the Bragg formalism expressed in angles and wavelengths and many experienced users of diffraction instruments continue to think naturally in this way. In this section, it is assumed that all elements have rectangular transmission profiles – so transmission is 100% up to some limit where it falls to 0%. The scattered beam is considered at the point $x_S$=0 which is assumed to be representative of the beam over the sample width of interest. Thus, the beam within the PS is assumed to be sufficiently wide that spatial effects can be ignored. The monochromator is assumed to be flat and oriented at an angle $\theta_M$ to the nominal incident beam direction. Local variations in crystal mosaic element orientation ($\eta$) or incident ray direction ($\gamma_0$) alter the effective Bragg angle to $\theta_M'=\theta_M+\eta-\gamma_0$ and the angle between the incident and scattered rays to $2\theta_M' = 2\theta_M+2\eta-\gamma_0$.

For a perfectly collimated "white" incident beam containing all wavelengths, a flat monochromator crystal of zero mosaic oriented at Bragg angle $\theta_M$ produces a monochromatic beam at scattering angle $2\theta_M$. The DuMond diagram for this situation shows a single point of intensity (figure 5a). A $\theta_M$-$2\theta_M$ scan of that monochromator (or alternatively, using a monochromator where the mosaic is $\eta_M = \pi$, *ie* a powder) produces the familiar curve of wavelength variation with scattering angle shown by the dotted line in figure 5b. If the monochromator mosaic has some finite value, intensity appears along a segment of that curve as shown by the solid curve segment in figure 5b.

If a "monochromator" with mosaic $\eta_M = \pi$ is now illuminated by a perfectly collimated white beam displaced by some small angle, $\gamma_0$, from the incident beam axis, the curve is also displaced by $\gamma_0$ parallel to the $2\theta$ axis. It follows that if the incident beam has some finite angular width, $\pm\alpha_0$, the curve is broadened by $\pm\alpha_0$ parallel to the $2\theta$ axis as in figure 5c. The effect of such a beam angular width on a monochromator with zero mosaic is illustrated by the short solid line segment.

Now consider a flat crystal monochromator of zero mosaic at orientation $\theta_M$ illuminated by a "white" beam with all divergences between glancing and normal reflection from the monochromator. Rays will be diffracted with incident Bragg angles varying from 0° (scattered angle $\theta_M$) to back scattering (scattered angle $\theta_M+90°$) with corresponding variation in wavelength as shown by the dashed line in figure 5d. The effect of finite monochromator mosaic is shown by the short solid line segment and this results in broadening the dashed curve as shown by the solid lines in figure 5d.

Combining the effects of finite mosaic and incident beam divergence gives an acceptance area on the DuMond Diagram as shown in figure 5e. This defines the $2\theta$-$\lambda$ range available in the beam following the monochromator. The effect of collimation between the monochromator and the sample, $\alpha_1$, is to limit the angular range of rays visible at the sample as shown in figure 5f.

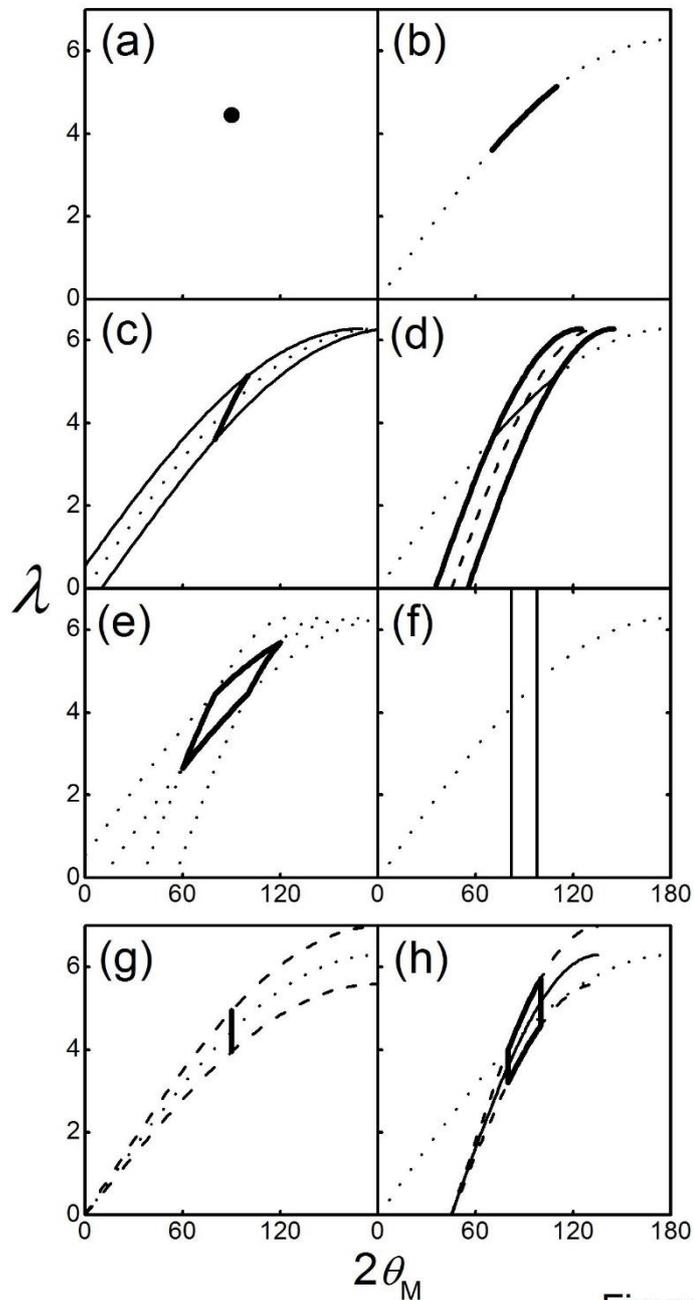

Figure 5

Figures 5c, d & f illustrate the transmission allowed by each element. The total transmission is all rays which can pass all three PS elements *ie* the product or superposition of these three DuMond Diagrams corresponding to the individual PS elements.

Now, consider a gradient crystal monochromator with zero mosaic but a range of *d*-spacings, described by a fractional variation $\mu_M$, such that $d_M(1-\mu_M) < d < d_M(1+\mu_M)$. A perfectly collimated incident white beam is scattered at a single angle but with a range of wavelengths as shown by the short vertical solid line in figure 5g. The dashed lines show how this wavelength range depends on Bragg angle, as would be seen in a $\theta_M$-$2\theta_M$ rocking scan. Notice that for a gradient monochromator the wavelength spread is larger at large $\theta_M$, *ie* large wavelengths ($|\Delta\lambda| \leq \lambda\mu_M$) in contrast to the reduced wavelength spread at large scattering angles and wavelengths seen with mosaic monochromators ($|\Delta\lambda| \leq \eta_M\lambda\cot\theta_M$). This suggests that in some applications gradient monochromators would have an advantage for short wavelength neutrons where $\theta_M$ is usually small leading to a large $\cot\theta_M$ and poor $\lambda$ resolution. Attempts have been made to construct gradient monochromator crystals for many years with limited success [*eg* 18]. It may be possible to manufacture composite gradient crystals from a stack of thin crystal plates each with slightly different concentration and hence *d*-spacing (made, for example, using a "micro pull down" furnace [19]). Elastically bending crystals induces a lattice spacing gradient so if necessary the stack of crystal plates could be bent slightly to make the *d*-spacing gradient continuous rather than discrete. The clever use of slabs of the comparatively cheap and readily available large crystals of near perfect silicon or germanium manufactured for the semiconductor industry as bent neutron monochromators is now relatively common. The AD formalism should apply to such monochromators, at least approximately. For such monochromators the degree of gradient induced is coupled to the monochromator curvature which is often adjusted when $\theta_M$ is changed.

Figure 5h shows the effect of a white beam with angular width $\alpha_0$ incident on a gradient monochromator. The solid lines outline the DuMond diagram for finite $\alpha_0$ while the dashed lines show the effect if $\alpha_0=\pi$. Because $\gamma_0 = -\gamma_1$ in scattering from a flat gradient monochromator, $\alpha_0$ has the same effect as $\alpha_1$.

The spreads in angle and wavelength for neutron primary spectrometers are typically quite small (of order 0.5° or 0.5%). Under these circumstances, the curvature of the lines bounding the DDs becomes insignificant and they can be represented accurately enough as straight lines. In this limit, starting with a perfectly collimated incident beam, $\alpha_0=0$, and a monochromator for which $\eta_M=\mu=0$:

- A spread $\pm\alpha_0$ in $\gamma_0$ broadens the 2D ($2\theta_M$, $\lambda$) DuMond diagram point by $\pm(1,\lambda\cot\theta_M)\alpha_0$

- Monochromator mosaic $\eta_M$ broadens the DuMond diagram point by $\pm(1, \frac{1}{2}\lambda\cot\theta_M)\eta_M$

- Monochromator *d*-spacing gradient $\pm\mu$ broadens the DuMond diagram point by $\pm(0, 1)\mu\lambda$

- Angular collimation in the scattered beam limits the DuMond diagram by lines at $2\theta_M\pm\alpha_1$

## 4. Acceptance diagrams for individual PS elements

The conventional primary spectrometer considered here is itself a diffraction instrument and the viewpoint most likely to be informative is a wave-vector space. The DuMond diagrams in section 3 can be equivalently presented as $AD_S$, the sample beam Acceptance Diagram drawn in an angle : wave-vector space. The AD plotted in a $\gamma$-$\Delta\kappa/\kappa$ space is just the reflection in the $\gamma$ axis of the DuMond diagram plotted in a $\gamma$-$\Delta\lambda/\lambda$ space.

This section develops expressions describing the AD limits in wave-vector space for different types of collimating elements and for curved as well as flat monochromator crystals. It is assumed that the monochromator is thin and that any curvature is continuous although in practice the curvature is usually achieved approximately using small oriented crystal segments. Figure 6a shows $AD_S$ and its component parts for individual PS elements corresponding to figs 5c, d & f. Figure 6b shows $AD_S$ for the gradient monochromator case corresponding to fig 5h.

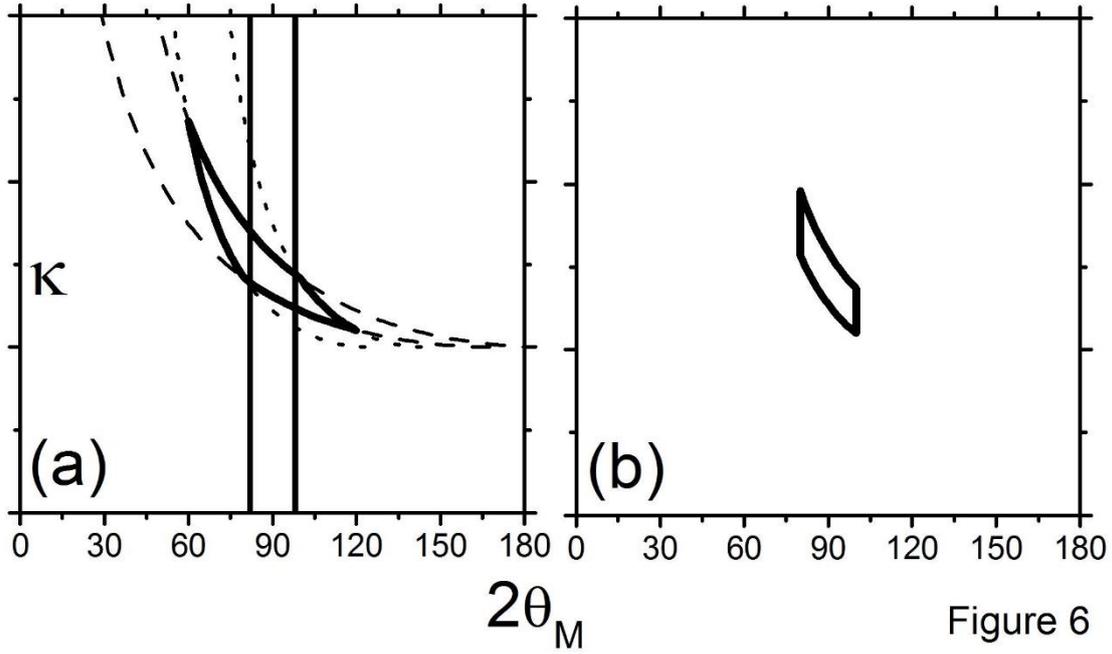

Figure 6

A small angular variation from the nominal beam direction at the sample, $\gamma_1$, equates to a small fractional variation in $\kappa_x$, the x-component of the wave-vector ($\gamma_1 \approx \sin\gamma_1 = \Delta\kappa_x/\kappa$). If $\gamma_1$ is small, then $\kappa_z \approx \kappa$ and thus $\Delta\kappa \approx \Delta\kappa_z$. Formally, $AD_S$ represents $\tau(\gamma_1,\kappa)$ in spherical polar coordinates or $\tau(\kappa_x,\kappa_z)$ in Cartesian coordinates. For mathematical convenience, the representation used here is $AD_S=\tau(\gamma_1, \Delta\kappa/\kappa) \cong \tau(\Delta\kappa_x/\kappa, \Delta\kappa_z/\kappa)$ which displays equivalent information. The beam is assumed to be sufficiently spatially uniform over the sample for any non-uniformity to have negligible effect on measurements (usually a necessary condition for sensible measurements) so spatial variations are largely ignored. Achieving this in practice usually simply requires "sufficiently large" PS dimensions.

For the case of rectangular profile collimators with a flat mosaic monochromator as considered in section 3 and assuming that all divergences are relatively small, $AD_S$ is bounded by three pairs of straight lines:

$$\Delta\kappa_Z/\kappa = -\tfrac{1}{2}cot\theta_M(\gamma_1 \pm \alpha_0) \quad \Delta\kappa_z/\kappa = -cot\theta_M(\gamma_1 \pm \eta_M) \quad \Delta\kappa_x/\kappa \approx \gamma_1 = \pm\alpha_1$$

For a gradient monochromator the AD is bounded by two pairs of lines:

$$\Delta\kappa_Z/\kappa = (-\gamma_1 \pm \mu_M) \quad \Delta\kappa_x/\kappa \approx \gamma_1 = \pm\alpha_1$$

Just as the sample position DD is the product of three component DD's, so $AD_S$ is the superposition or product of three AD's corresponding to the effects of the $\alpha_0$ collimation, the monochromator and the $\alpha_1$ collimation denoted $AD_{\alpha0}$, $AD_{Mono}$ and $AD_{\alpha1}$.

$$AD_{xS=0} = AD_{\alpha0} \bullet AD_{Mono} \bullet AD_{\alpha1}$$

It is simpler and clearer to consider $AD_S$ as this product than as a single whole.

$AD_{\alpha0}$ and $AD_{Mono}$ depend on scattering at the monochromator which allows a range in wave-vector given by equation 2. The derivations are conducted here only for beamtubes delivering rectangular angular profiles but non rectangular angular profiles can be superimposed later as a modulation of $\tau$ along well defined directions. Setting $L_0=\infty$ describes the guide and Soller collimator cases.

### 4.1 $AD_{\alpha 0}$ for a mosaic monochromator

$AD_{\alpha 0}$ includes all neutron rays which can reach $x_S=0$ regardless of the values of $R_{MH}$ and $\eta_M$. $AD_{\alpha 0}$ is derived by assuming that $\eta_M$ is large enough that all incident rays allowed by $\alpha_0$ can be scattered at the monochromator. Consider an open beamtube between V and M with a fully illuminated (virtual) source so that rays reach all parts of the monochromator (the "global" incident divergence is at least $\pm W_M \sin\theta_M/L_0$) but at each point $x_M$ on the monochromator, the "local" angular spread is $\alpha_0 \approx \pm W_V/L_0$ which may be much smaller. The centreline of $AD_{\alpha 0}$ is calculated by setting $\alpha_0=0$; for a beamtube this is achieved by setting the source width $2W_V$ to zero. Then, $\gamma_{0\_Mean} \approx x_{M0}\sin\theta_M/L_0$ and applying equations 3c and 5 yields

$$\frac{\Delta\kappa}{\kappa} \approx -\frac{1}{2}\gamma_1\left(1-\frac{L_1}{L_0}\right)\cot\theta_M \tag{Eq. 6}$$

This describes a line of intensity in $AD_S$. Introducing a finite $\alpha_0$ broadens this line in $AD_S$ along a direction calculated by now setting $\eta_M = 0$. Applying equations 2, 3a and 4 yields

$$\frac{\Delta\kappa}{\kappa} \approx -\gamma_1\left(1-\frac{L_1}{R_{MH}\sin\theta_M}\right)\cot\theta_M \tag{Eq. 7}$$

Equations 3a, 3b, 4 and 5 show that $\gamma_0 \approx 2\xi - \gamma_1 \approx \gamma_1\left(\frac{2L_1}{R_{MH}\sin\theta_M}-1\right)$ and $\left|\gamma_0 - \frac{x_{M0}\sin\theta_M}{L_0}\right| < \alpha_0$ so that the limits of this broadening line are set by the value of $\alpha_0$ as $|\gamma_1| < \alpha_0/\left(1+\frac{L_1}{L_0}-\frac{2L_1}{R_{MH}\sin\theta_M}\right)$. Calculating $\Delta\kappa/\kappa$ at the end of this broadening line and subtracting the value of $\Delta\kappa/\kappa$ at the centreline there gives a $\Delta\kappa/\kappa$ broadening of $\pm\frac{1}{2}\alpha_0\cot\theta_M$ regardless of the values of $L_0$ or $R_{MH}$. Thus, $AD_{\alpha 0}$ is the region bounded by the lines

$$\frac{\Delta\kappa}{\kappa} \approx -\gamma_1\left(1-\frac{L_1}{L_0}\right)\cot\theta_M \pm \frac{1}{2}\alpha_0\cot\theta_M \tag{Eq. 8}$$

The $\gamma_1$ extent of $AD_{\alpha 0}$ is limited by the values of $\eta_M$ and $\alpha_1$.

### 4.2 $AD_{\alpha 0}$ for a gradient monochromator

For a gradient monochromator, $\eta_M=0$, so a ray can only scatter if $\gamma_1-\xi = \xi - \gamma_0$ where $\xi=x_{M0}/R_{MH}$. $\gamma_0$ lies in the range $x_{M0}\sin\theta_M/L_0 \pm \alpha_0$ with $\gamma_1=x_{M0}\sin\theta_M/L_1$. The effect of $\alpha_0$ is simply to restrict the range of $\gamma_1$ with limits $|\gamma_1| < \alpha_0/\left(1+\frac{L_1}{L_0}-\frac{2L_1}{R_{MH}\sin\theta_M}\right)$ as before. Note that this also affects the sample beam spatial size which is limited to $\pm\alpha_0 L_1$ if the monochromator is fully focussed from source to sample. For a flat monochromator the sample beam width is limited by the source and monochromator widths and the allowed beam divergence.

### 4.3 $AD_{Mono}$

$AD_{Mono}$ includes all neutron rays which can reach $x_S=0$ regardless of the values of $L_0$ or $\alpha_0$. Assuming that $\alpha_0$ is unrestricted and conducting the derivation as before for a curved monochromator on a beamtube, the $AD_{Mono}$ centreline is calculated by setting $\eta_M = 0$ when $\xi \approx x_{M0}/R_{MH}$. Applying equations 3a and 5 yields

$$\frac{\Delta\kappa}{\kappa} \approx -\gamma_1\left(1-\frac{L_1}{R_{MH}\sin\theta_M}\right)\cot\theta_M \tag{Eq. 9}$$

A finite $\eta_M$ broadens this line along $\frac{\Delta\kappa}{\kappa} \approx -\frac{1}{2}(\gamma_1 - \gamma_0)\cot\theta_M \approx -\frac{1}{2}\gamma_1\left(1 - \frac{L_1}{L_0}\right)\cot\theta_M$ calculated by setting $\alpha_0 = 0$ so that at $x_{M0}$, $\gamma_0 \approx \frac{x_{M0}\sin\theta_M}{L_0} \approx \gamma_1\frac{L_1}{L_0}$. The limits to this broadening line are set by $\eta_M$ (or if not, by $\alpha_1$ or $W_M$). The equality $\xi = \frac{1}{2}(\gamma_0 + \gamma_1) = \frac{x_{M0}}{R_{MH}} + \eta = \frac{1}{2}\gamma_1\left(1 + \frac{L_1}{L_0}\right) = \frac{\gamma_1 L_1}{R_{MH}\sin\theta_M} + \eta$ shows that $\gamma_1$ lies in the range $|\gamma_1| < 2\eta_M/\left(1 + \frac{L_1}{L_0} - \frac{2L_1}{R_{MH}\sin\theta_M}\right)$. Calculating $\Delta\kappa/\kappa$ at the end of this broadening line and subtracting the centreline there gives a $\Delta\kappa/\kappa$ broadening of $\pm\eta_M\cot\theta_M$ regardless of $R_{MH}$ or $L_0$. A curved monochromator can be regarded as having small mosaic segments selected in some organised way from the $AD_{Mono}$ of some larger mosaic flat monochromator.

A monochromator $d$-spacing gradient broadens the line $\frac{\Delta\kappa}{\kappa} \approx -\gamma_1\left(1 - \frac{L_1}{R_{MH}\sin\theta_M}\right)\cot\theta_M$ by $\frac{\Delta\kappa}{\kappa} \approx \pm\mu$ with the $\gamma_1$ limits given above.

Figure 7a illustrates the broadening effect on $AD_{\alpha0}$ of $\alpha_0 = 30'$ with the centreline of $AD_{\alpha0}$ indicated by the dashed line (here $\theta_M = 45°$). Figure 7b shows the broadening effect on $AD_{Mono}$ of $\eta_M = 10'$ with the $AD_{Mono}$ centreline shown by the dashed line (again, $\theta_M = 45°$).

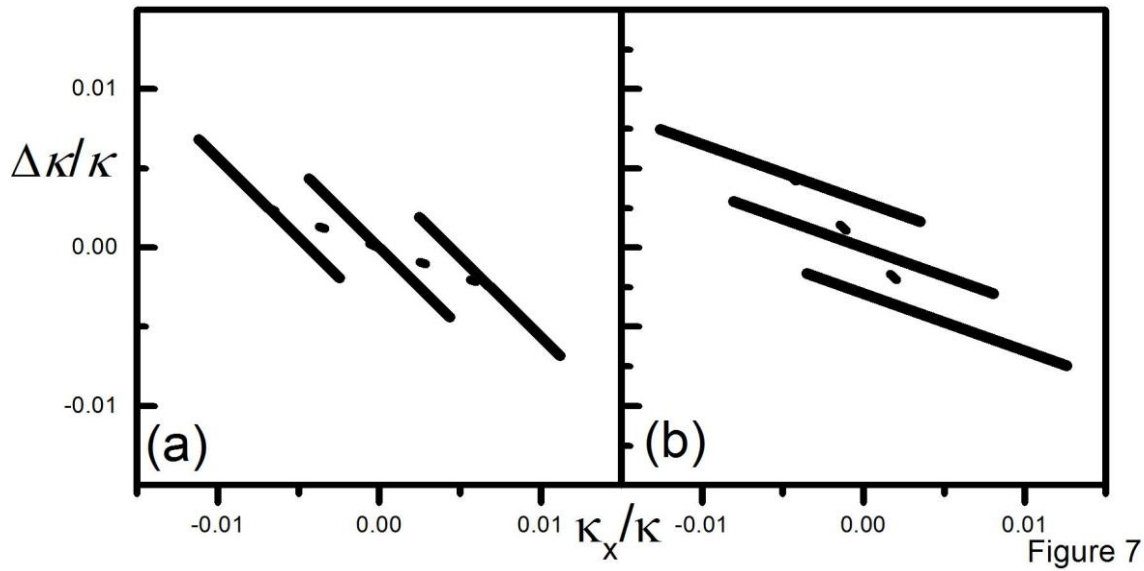

Figure 7

Figure 8a shows a McStas simulation of the sample position beam from a PS using a flat monochromator with a very small $\alpha_0$ (3') and $\eta_M = 20'$. This gives an $AD_S$ which follows the $AD_{Mono}$ broadening line. Figure 8b shows a simulation of beam with a very small $\eta_M$ (3') and $\alpha_0$=30' showing the $AD_{\alpha 0}$ broadening line.

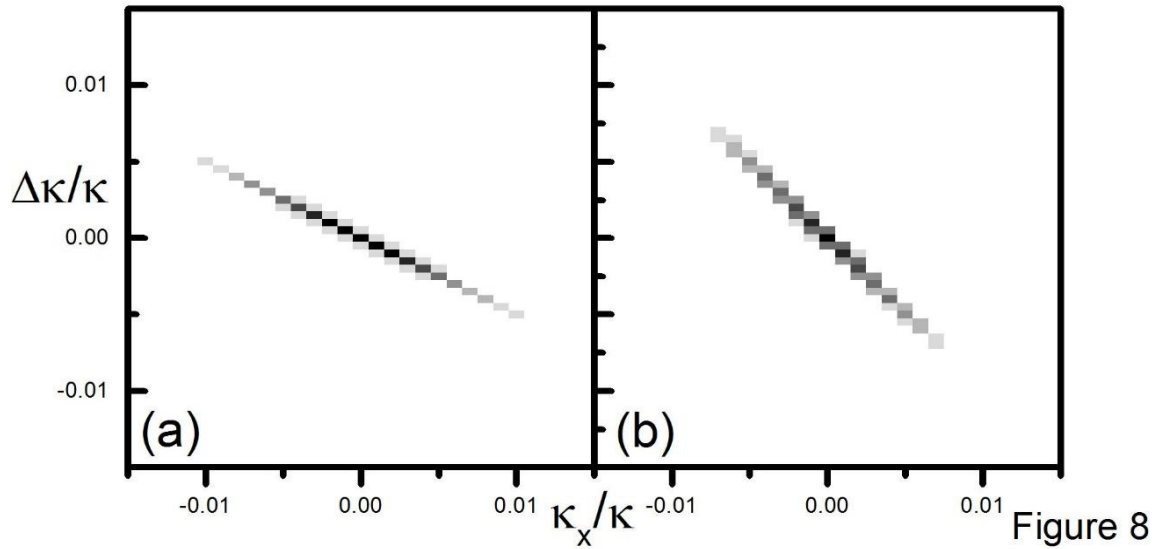

Figure 8

### 4.4 $AD_{\alpha 1}$ - Collimator between the monochromator and sample

Beam collimation between the monochromator and sample plainly has no effect on wave-vector and simply restricts the sample beam's angular divergence distribution. $AD_{\alpha 1}$ consists of equi-transmission contours parallel to the $\Delta\kappa/\kappa$ axis with the profile, $\tau(\gamma_1)$, dependent on the type of collimation. At $x_S$=0 the angular distribution is triangular for a Soller collimator and rectangular for an ideal guide with FW = $\pm\alpha_1$. For a beamtube or converging Radial Soller collimator, the angular distribution at $x_S$=0 is rectangular with FW = $\pm\alpha_1$ where $\alpha_1 \approx W_M \sin\theta_M/L_1$. The illuminated monochromator width sets an upper limit of $\pm W_M \sin\theta_M/L_1$ on angular width for all $\alpha_1$ collimators.

In this discussion of AD shapes, no consideration has been taken yet of the absolute value of the "transmission". The transmission at each point in $AD_S$ should be 100%, multiplied by any losses due to collimator transmission, monochromator reflectivity, air scattering or absorption by any windows in the beam. The transmission must be modulated along the individual AD axes by consideration of any transmission profile due to $\alpha_0$ or $\alpha_1$ collimation or to monochromator mosaic. The total beam flux at $x_S$=0 (clearly ignoring vertical divergence effects) is the source flux multiplied by the integral of $\tau$ over $AD_S$. Thus, the intensity depends on the Acceptance Diagram area so that a larger AD, corresponding to a beam with large angular and wave-vector spreads, represents a higher intensity (with correspondingly reduced resolution). The instrument count rate will be proportional to this intensity multiplied by the sample area.

### 5. Sample position Acceptance Diagram in the Gaussian approximation

This section discusses the product $AD_S$ for "Gaussian" elements, an approximation often applied to triangular transmission functions. Considerable work and many publications have been devoted to deriving the full resolution function for many types of neutron scattering instruments including those using HFMs. This section is not intended as any replacement for any existing formalism (*eg* 4,5,15,16) but simply as a bridge from the pictorial view described in sections 2, 3 and 4 to the significant results sought in section 6.

To generate (approximately) Gaussian profiles in each PS component, consider a PS using a Radial Soller collimator for $\alpha_0$, a Gaussian mosaic HFM and a conventional Soller collimator for $\alpha_1$. The individual transmission functions are then

$$\tau_{\alpha 0}\left(\gamma_1, \frac{\Delta\kappa}{\kappa}\right) = exp - \frac{8ln2}{2}\left\{\left(\frac{\left(\frac{\Delta\kappa}{\kappa}+\frac{1}{2}\gamma_1\left(1-\frac{L_1}{L_0}\right)cot\theta_M\right)}{\frac{1}{2}\alpha_0 cot\theta_M}\right)^2\right\} \quad \text{(Eq. 10a)}$$

$$\tau_{Mono}\left(\gamma_1, \frac{\Delta\kappa}{\kappa}\right) = exp - \frac{8ln2}{2}\left\{\left(\frac{\left(\frac{\Delta\kappa}{\kappa}+\gamma_1\left(1-\frac{L_1}{R_{MH}sin\theta_M}\right)cot\theta_M\right)}{\eta_M cot\theta_M}\right)^2\right\} \quad \text{(Eq. 10b)}$$

$$\tau_{\alpha 1}\left(\gamma_1, \frac{\Delta\kappa}{\kappa}\right) = exp - \frac{8ln2}{2}\left\{\left(\frac{\gamma_1}{\alpha_1}\right)^2\right\} \quad \text{(Eq. 10c)}$$

with the 8ln2 factors in each Gaussian allowing $\alpha_0$, $\eta_M$ and $\alpha_1$ to be written as FWHM. $AD_S$, the total AD at the sample centre, $x_S$=0, is the product of eqs 10a, 10b and 10c and is

$$\tau_{xS=0}\left(\gamma_1, \frac{\Delta\kappa}{\kappa}\right) = exp - \frac{8ln2}{2}\left\{A\gamma_1^2 + B\gamma_1\frac{\Delta\kappa}{\kappa} + C\left(\frac{\Delta\kappa}{\kappa}\right)^2\right\} \quad \text{(Eq. 11a)}$$

where

$$A = \left(\left(1 - \frac{L_1}{L_0}\right)^2 \alpha_0^{-2} + \left(1 - \frac{L_1}{R_{MH}\sin\theta_M}\right)^2 \eta_M^{-2} + \alpha_1^{-2}\right)$$
$$B = \left(4\left(1 - \frac{L_1}{L_0}\right)\alpha_0^{-2} + 2\left(1 - \frac{L_1}{R_{MH}\sin\theta_M}\right)\eta_M^{-2}\right)\tan\theta_M \quad \text{(Eq. 11b)}$$
$$C = (4\alpha_0^{-2} + \eta_M^{-2})\tan^2\theta_M$$

Eq. 11a should include transmission terms to describe any beam losses in the PS but these are all assumed to be 100% here. Fixing $\tau$ in eq. 11a to some value gives an elliptic contour of constant transmission probability. For the significant contour at $\tau=\frac{1}{2}$, $y = \sqrt{C - (4AC - B^2)x^2} - \frac{Bx}{2C}$ and this contour

- intersects the $\Delta\kappa/\kappa$ axis at $\quad\quad\quad\quad\quad\quad\quad\quad\quad\quad\quad\quad\quad\quad\quad \pm \frac{1}{2}\, C^{-1/2}$

- intersects the $\gamma_1$ axis at $\quad\quad\quad\quad\quad\quad\quad\quad\quad\quad\quad\quad\quad\quad\quad\quad\quad \pm \frac{1}{2}\, A^{-1/2}$

- has maximum extent in the $\gamma_1$ direction of $\quad\quad\quad\quad \gamma_{Max} = \pm\sqrt{C/(4AC - B^2)}$

- has maximum extent in the $\Delta\kappa/\kappa$ direction of $\quad\quad \left(\frac{\Delta\kappa}{\kappa}\right)_{Max} = \pm\sqrt{A/(4AC - B^2)}$

- has angle between the $\gamma_1$-axis and the ellipse axis of $\frac{1}{2}\tan^{-1}(B/(C-A))$

The known and well tested expressions for CW PDs and TAS resolution using Soller collimators and flat monochromators can be recovered from Eq 11 by setting $L_0 = R_{MH} = \infty$ when

$$A = (\alpha_0^{-2} + \eta_M^{-2} + \alpha_1^{-2})$$
$$B = (4\alpha_0^{-2} + 2\eta_M^{-2})\tan\theta_M \quad \text{(Eq. 11c)}$$
$$C = (4\alpha_0^{-2} + \eta_M^{-2})\tan^2\theta_M$$

and this should give confidence in this extension of the AD description to non Gaussian collimators and HFMs. Eq. 11 for the PS AD should be directly transferrable to the known expressions describing the resolution of CW PDs and TAS permitting the inclusion of HFMs in resolution calculations.

Note that altering $L_1/L_0$ by varying either $L_0$ or $L_1$ for a radial Soller collimator $\alpha_0$ shears $AD_{\alpha 0}$ parallel to the $\Delta\kappa/\kappa$ axis. Altering $R_{MH}$ shears $AD_{Mono}$ parallel to the $\Delta\kappa/\kappa$ axis but does not affect the area of $AD_{Mono}$. Adjusting either $L_1/L_0$ or $R_{MH}$ affects the overlap between $AD_{\alpha 0}$ and $AD_{Mono}$ showing that increased intensity at the sample from horizontal monochromator curvature results from better matching $AD_{Mono}$ to $AD_{\alpha 0}$ giving access to a larger $\gamma_1$ range. That effect is maximised when the slopes of $AD_{Mono}$ and $AD_{\alpha 0}$ are exactly matched which corresponds to the well known source to sample "focussing condition" for a HFM when

$$\frac{1}{2}\left(1 - \frac{L_1}{L_0}\right) = \left(1 - \frac{L_1}{R_{MH}\sin\theta_M}\right) \quad \text{and thus} \quad R_{MH} = \frac{2L_1 L_0}{(L_0 + L_1)\sin\theta_M} \quad \text{(Eq. 12)}$$

If $\alpha_0$ is a guide ($L_0 = \infty$), then $R_{MH} = \frac{2L_1}{\sin\theta_M}$ and if $L_0 = L_1$, (the "monochromatic" focussing condition), then $R_{MH} = \frac{L_1}{\sin\theta_M}$. If the monochromator curvature takes the eq. 12 value,

$$A = \left(\frac{1}{4}(4\alpha_0^{-2} + \eta_M^{-2})\left(1 - \frac{L_1}{L_0}\right)^2 + \alpha_1^{-2}\right) \quad B = (4\alpha_0^{-2} + \eta_M^{-2})\left(1 - \frac{L_1}{L_0}\right)\tan\theta_M \quad C = (4\alpha_0^{-2} + \eta_M^{-2})\tan^2\theta_M$$

and

$$\tau_{xS=0} = exp-\frac{8ln2}{2}\left\{(4\alpha_0^{-2} + \eta_M^{-2})\left(tan\theta_M \frac{\Delta\kappa}{\kappa} + \frac{1}{2}\left(1 - \frac{L_1}{L_0}\right)\gamma_1\right)^2 + \alpha_1^{-2}\gamma_1^2\right\} \quad \text{(Eq. 13)}$$

The equations suggest that if the monochromator curvature is convex rather than concave, then the slope of $AD_{Mono}$ increases and, presumably, matching this would require a convergent rather than divergent beam before the monochromator.

### 6. Relationship between instrument parameters ($\alpha_0, \alpha_1, \eta_M, \theta_M, L_1, L_0, R_{MH}$) and (A, B, C)

This section discusses the range of PS parameters which can be used to deliver some chosen beam characteristics.

Within a Gaussian approximation, the values of (A,B,C) (eq. 11b) can be used to draw elliptical contours of constant transmission probability in $AD_S$ as described in a $(\gamma_1, \Delta\kappa/\kappa)$ or $(\Delta\kappa_x/\kappa, \Delta\kappa_z/\kappa)$ space. As far as scattering by the sample is concerned, any beam of a given $\kappa$ with identical values for the variables (A,B,C) has an identical effect. The description would be more useful in discussing instrument resolution if it used parameters simply related to the observed scattering. It is useful to consider a "delta function scatterer" sample which transforms incident neutrons by some fixed $\Delta(\underline{\kappa},\omega)$. Such a sample represents a Bragg peak in the elastic scattering case (where $\omega$=0). For Bragg peak sample scattering, equation 2 can be adjusted to become $\Delta\theta_S \approx -(\Delta\kappa/\kappa)tan\theta_S$ showing that scattering shears $AD_S$ parallel to the $\gamma$ axis. A useful parameter set to describe $AD_S$, at least for CW PDs, proves to be $(u, \psi_{AD}, \alpha_{In})$ as illustrated in figure 9. Here $\alpha_{In}$ is the beam angular width (FWHM) at the nominal wave-vector measured at the $\gamma_1$ axis. $u$ is the maximum extent (FWHM) of the $\tau=\frac{1}{2}$ elliptical contour in the $\Delta\kappa/\kappa$ direction. $\psi_{AD}$ is the angle between the $\Delta\kappa/\kappa$ axis and the line joining the origin to this point of maximum $\Delta\kappa/\kappa$ extent. Note that $\psi_{AD}$ is negative for positive $\theta_M$. The parameter $u$ is equal to $\sqrt{U}$ where $U$ is the first of the parameter set (U,V,W) commonly used in Rietveld analysis of powder diffraction patterns. $u$ describes the degree of peak broadening arising from wavelength spread in the beam. In the Gaussian model, $\psi_{AD}$ determines the scattering angle on a PD at which the peak width is smallest. These parameters can be seen to be related to the notion that on a PD the in-plane contribution to peak broadening arises from a combination of wavelength spread and angular spread in the incident beam. It is straightforward to show that

$$u = 2\sqrt{\frac{4A}{4AC-B^2}}$$
$$\Psi_{AD} = atan\left(\frac{-B}{2A}\right) \quad \text{and} \quad \begin{aligned} A &= \alpha_{In}^{-2} \\ B &= -2\alpha_{In}^{-2}tan\Psi_{AD} \\ C &= 4u^{-2} + \alpha_{In}^{-2}tan^2\Psi_{AD} \end{aligned} \quad \text{(Eq. 14)}$$
$$\alpha_{In} = \frac{1}{\sqrt{A}}$$

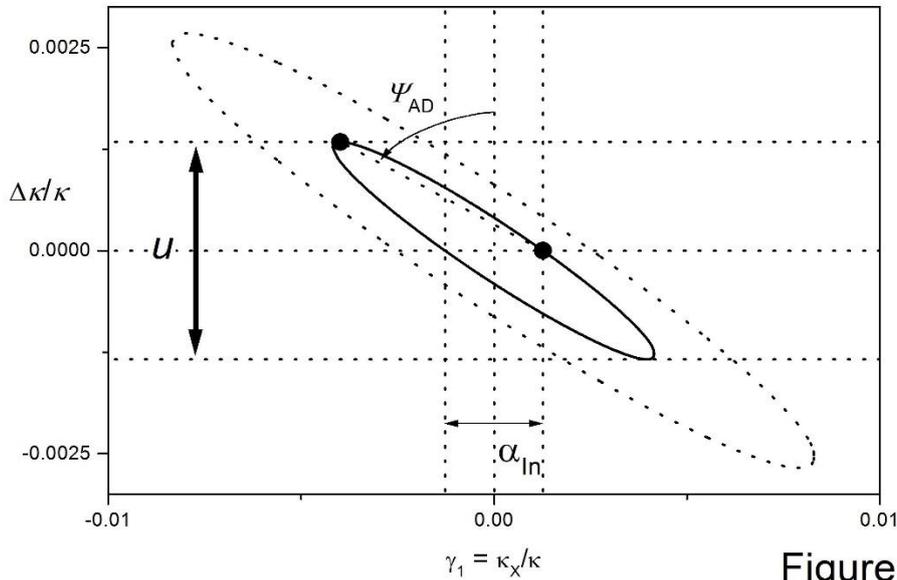

Figure 9

It has been shown [20] that many values of ($\alpha_0$, $\alpha_1$, $\eta_M$, $\theta_M$) can deliver identical resolution on CW PDs within the Gaussian approximation for a Soller – flat mosaic monochromator – Soller PS. Such parameter sets all have the same detector collimation so this parameter multiplicity is plainly associated only with the PS. If the desired beam character is known, then, starting from equation 11b and the known values for (A,B,C) or ($u,\psi_{AD},\alpha_{In}$), parameter values can be deduced which deliver that beam in the more general case discussed here which allows for HFMs and open beamtubes. The mathematics defines 3 functions of 7 parameters and deducing the relations requires fixing 4 of the parameters. There are many combinations of parameters which could be fixed but only three cases are treated here. The first case considers conventional Soller $\alpha_0$ and $\alpha_1$ collimators with a flat mosaic monochromator (fixing $L_1$, $L_0$, $R_{MH}$). The second case considers conventional Soller or guide $\alpha_0$ and $\alpha_1$ collimators with a mosaic monochromator curved to focus from source to sample (with $L_0=\infty$ and $L_1$ fixed). The third case uses a radial Soller for $\alpha_0$, a curved mosaic monochromator and a conventional Soller collimator for $\alpha_1$ with the axes of $AD_{\alpha 0}$ and $AD_{Mono}$ aligned (fixing $L_1$, coupling $R_{MH}$ and $L_0$ and coupling $\alpha_0$ and $\eta_M$). In cases 1 & 3, $\theta_M$ is used as a variable to display the range of values possible.

### 6.1 Soller collimator $\alpha_0$ and $\alpha_1$ with flat mosaic crystal monochromator

The case of a conventional primary spectrometer using Soller collimators for $\alpha_0$ and $\alpha_1$ and a flat mosaic monochromator crystal corresponds to setting $L_0=R_{MH}=\infty$. Then it is straightforward starting from eq. 11b and fixing $\theta_M$ to show that

$$\begin{aligned} A &= \left(\alpha_0^{-2} + \eta_M^{-2} + \alpha_1^{-2}\right) \\ B &= \left(4\alpha_0^{-2} + 2\eta_M^{-2}\right)tan\theta_M \\ C &= \left(4\alpha_0^{-2} + \eta_M^{-2}\right)tan^2\theta_M \end{aligned} \quad \text{or} \quad \begin{aligned} u &= 2cot\theta_M \sqrt{\frac{\left(\alpha_0^2\alpha_1^2+\alpha_0^2\eta_M^2+\eta_M^2\alpha_1^2\right)}{\left(\alpha_0^2+4\eta_M^2+\alpha_1^2\right)}} \\ \Psi_{AD} &= atan\left(\frac{-\left(\alpha_0^2\alpha_1^2+2\eta_M^2\alpha_1^2\right)}{\left(\alpha_0^2\alpha_1^2+\alpha_0^2\eta_M^2+\eta_M^2\alpha_1^2\right)}tan\theta_M\right) \\ \alpha_{In} &= \left(\alpha_0^{-2} + \eta_M^{-2} + \alpha_1^{-2}\right)^{-1/2} \end{aligned}$$

Applying equation 14 shows that

$$\begin{aligned} \alpha_0^{-2} &= \frac{1}{2}Ccot^2\theta_M - \frac{1}{4}Bcot\theta_M &&= \frac{1}{2}\alpha_{In}^{-2}tan\Psi_{AD}cot\theta_M\{1+tan\Psi_{AD}cot\theta_M\} + 2u^{-2}cot^2\theta_M \\ \eta_M^{-2} &= Bcot\theta_M - Ccot^2\theta_M &&= -\alpha_{In}^{-2}tan\Psi_{AD}cot\theta_M\{2+tan\Psi_{AD}cot\theta_M\} - 4u^{-2}cot^2\theta_M \\ \alpha_1^{-2} &= A - \frac{3}{4}Bcot\theta_M + \frac{1}{2}Ccot^2\theta_M &&= \frac{1}{2}\alpha_{In}^{-2}\{2+tan\Psi_{AD}cot\theta_M\}\{1+tan\Psi_{AD}cot\theta_M\} + 2u^{-2}cot^2\theta_M \end{aligned}$$

Instrument parameters can be found to deliver the sample position beam described by {A,B,C} or {$u,\psi_{AD},\alpha_{In}$} for a range of $\theta_M$ values. That range can be found using the requirement that $\alpha_0^{-2}$, $\eta_M^{-2}$ and $\alpha_1^{-2}$ all be positive; so

$$B/C > cot\theta_M > B/2C \quad \text{or} \quad \frac{-\alpha_{In}^{-2}tan\Psi_{AD}}{[\alpha_{In}^{-2}tan^2\Psi_{AD}+4u^{-2}]} < cot\theta_M < \frac{-2\alpha_{In}^{-2}tan\Psi_{AD}}{[\alpha_{In}^{-2}tan^2\Psi_{AD}+4u^{-2}]}.$$

The allowed $\theta_M$ range is often continuous but sometimes has a disallowed mid-region as is readily calculable as that part where $\alpha_1^{-2}$ would negative. One approach to choosing suitable values for instrument variables is to calculate and plot values for $\alpha_0$, $\eta_M$ and $\alpha_1$ over the allowed range of $\theta_M$ and then select a $\theta_M$ value for which the set of variables is technically convenient. Figures 10a and 10b show examples of the allowed values.

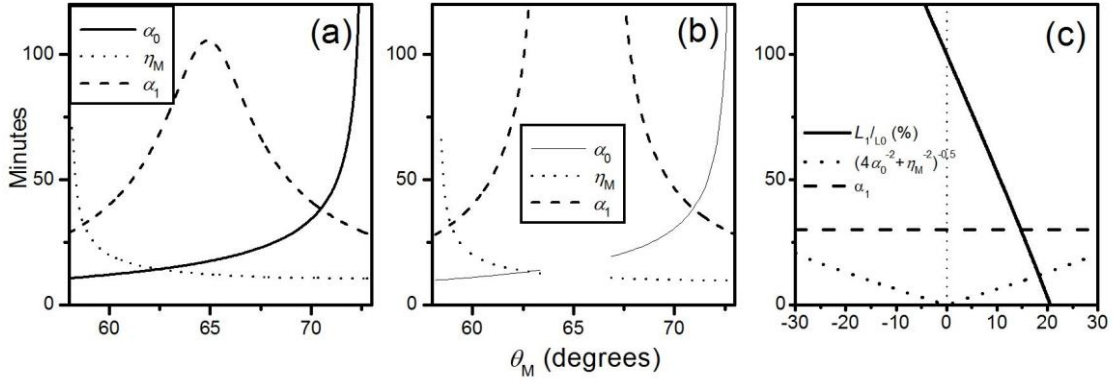

Figure 10

## 6.2 Soller collimators or guides using a focussed monochromator to align $AD_{\alpha 0}$ and $AD_{Mono}$

It is possible to curve the monochromator in the scattering plane and thus increase the alignment of $AD_{\alpha 0}$ and $AD_{Mono}$ which should improve the PS transmission. The case of a conventional primary spectrometer using Soller or guide collimators for $\alpha_0$ and $\alpha_1$ and a curved mosaic monochromator focussing to the sample corresponds to setting $L_0 = \infty$ and $R_{MH} = \frac{2L_1}{sin\theta_M}$. Then it is straightforward starting from eq. 11b and fixing $\theta_M$ to show that

$$A = \tfrac{1}{4}(4\alpha_0^{-2} + \eta_M^{-2}) + \alpha_1^{-2}$$
$$B = (4\alpha_0^{-2} + \eta_M^{-2})tan\theta_M$$
$$C = (4\alpha_0^{-2} + \eta_M^{-2})tan^2\theta_M$$

or

$$u = cot\theta_M \sqrt{\frac{(\alpha_0^2\alpha_1^2 + 4\alpha_0^2\eta_M^2 + 4\eta_M^2\alpha_1^2)}{(4\eta_M^2 + \alpha_0^2)}}$$
$$\Psi_{AD} = atan\left(\frac{-4(4\alpha_0^2\eta_M^2 + \alpha_0^2\alpha_1^2)}{(4\alpha_1^2\eta_M^2 + \alpha_0^2\alpha_1^2) + \alpha_0^2\eta_M^2} tan\theta_M\right)$$
$$\alpha_{In} = \left(\tfrac{1}{4}(4\alpha_0^{-2} + \eta_M^{-2}) + \alpha_1^{-2}\right)^{-1/2}$$

It follows that

$$(4\alpha_0^{-2} + \eta_M^{-2}) = B^2/C = 4u^2 tan^2\Psi_{AD}/(4\alpha_{In}^4 + u^2\alpha_{In}^2 tan^2\Psi_{AD})$$
$$\alpha_1^{-2} = A - \tfrac{1}{4}B^2/C = \alpha_{In}^{-2} - u^2 tan^2\Psi_{AD}/(4\alpha_{In}^4 + u^2\alpha_{In}^2 tan^2\Psi_{AD})$$
$$\theta_M = atan\left(\tfrac{C}{B}\right) = -atan\left(\frac{4u^{-2} + \alpha_{In}^{-2} tan^2\Psi_{AD}}{2\alpha_{In}^{-2} tan\Psi_{AD}}\right)$$

Here there is only one allowed value for $\theta_M$ but multiple values of $\alpha_0$ and $\eta_M$ so it is possible to choose $\alpha_0$ to match some preferred $\eta_M$ value.

If using Soller collimators, the distributions $\tau(\gamma_0)$ and $\tau(\gamma_1)$ are triangular. If using an ideal guide before the monochromator, $\tau(\gamma_0)$ becomes rectangular. It is possible to use a slit defining the beam width either just before or just after the monochromator to set the value for $\alpha_1$ and give a rectangular distribution.

Choosing rectangular distributions should increase transmission for a given resolution. In this case, the centre of $\tau(\gamma_1)$ varies with position at the sample; which may or may not have a significant effect on scans.

### 6.3 Focussed monochromator with $AD_{\alpha 0}$ and $AD_{Mono}$ aligned

For a PS using a radial Soller $\alpha_0$, curved mosaic monochromator and conventional Soller $\alpha_1$, matching the $AD_{\alpha 0}$ and $AD_{Mono}$ slopes and fixing $\theta_M$ yields

$$A = \left(\frac{1}{4}(4\alpha_0^{-2} + \eta_M^{-2})\left(1 - \frac{L_1}{L_0}\right)^2 + \alpha_1^{-2}\right)$$
$$B = (4\alpha_0^{-2} + \eta_M^{-2})\left(1 - \frac{L_1}{L_0}\right)\tan\theta_M$$
$$C = (4\alpha_0^{-2} + \eta_M^{-2})\tan^2\theta_M$$

or

$$u = 2\cot\theta_M\sqrt{\frac{1}{4}\left(1 - \frac{L_1}{L_0}\right)^2 \alpha_1^2 + (4\alpha_0^{-2} + \eta_M^{-2})^{-1}}$$
$$\cot\Psi_{AD} = -\cot\theta_M\left(\frac{1}{2}\left(1 - \frac{L_1}{L_0}\right) + \frac{2\alpha_1^{-2}}{(4\alpha_0^{-2}+\eta_M^{-2})\left(1-\frac{L_1}{L_0}\right)}\right)$$
$$\alpha_{In} = \left(\frac{1}{4}(4\alpha_0^{-2} + \eta_M^{-2})\left(1 - \frac{L_1}{L_0}\right)^2 + \alpha_1^{-2}\right)^{-1/2}$$

It follows then that

$$(4\alpha_0^{-2} + \eta_M^{-2}) = C\cot^2\theta_M = (4u^{-2} + \alpha_{In}^{-2}\tan^2\Psi_{AD})\cot^2\theta_M$$
$$\alpha_1^{-2} = A - \frac{1}{4}B^2/C = 4\alpha_{In}^{-2}u^{-2}/(4u^{-2} + \alpha_{In}^{-2}\tan^2\Psi_{AD})$$
$$\frac{L_1}{L_0} = 1 - B\tan\theta_M/C = 1 + 2\alpha_{In}^{-2}\tan\Psi_{AD}\tan\theta_M/(4u^{-2} + \alpha_{In}^{-2}\tan^2\Psi_{AD})$$

In this case, $\alpha_1$ always takes the same value for given ($A$,$B$,$C$) regardless of the value of $\theta_M$. The sum $(4\alpha_0^{-2}+\eta_M^{-2})$ is the significant variable so the individual values of $\alpha_0$ and $\eta_M$ can be varied freely as long as this sum remains constant. Figure 10c shows an example of the allowed values for a single choice of ($A$,$B$,$C$). The limits to $\theta_M$ are set by the requirement that $L_1/L_0$ be positive, so $\tan\theta_M < C/B$. Note that solutions exist for a reversed sign of $\theta_M$ when $L_0 < L_1$.

It is desirable to choose beam elements to deliver rectangular transmission profiles in $AD_S$ to increase transmission for given instrument resolution. This can be done using open beamtubes between virtual source and monochromator where the virtual source width sets $\alpha_0$ and between monochromator and sample where the effective monochromator width, which could be controlled by a slit just before or just after the monochromator, sets $\alpha_1$. To achieve the same resolution variance for a PS using rectangular elements the HW values for $\alpha_0$, $\eta_M$ and $\alpha_1$ should be approximately $\sqrt{2}$ times smaller than the Gaussian FWHMs. The relations presented above can be regarded as indicative of appropriate values for rectangular profile elements but should be used with some care. For example, for an $AD_S$ with rectangular transmission profiles, the proper value for $\psi_{AD}$ for a CW PD is probably the angle from the origin to the upper left vertex of $AD_S$, *ie* $\tan\Psi_{AD} = \alpha_1/\left(\frac{1}{2}\left(1 - \frac{L_1}{L_0}\right)\alpha_1 + \beta\right)$ where $\beta$ is the smaller of $\alpha_0/2$ and $\eta_M$.

### 6.4 Some examples of equivalent primary spectrometers

While the mathematics is clear, some of the conclusions may be surprising. To provide independent support for the accuracy of the mathematics, a series of McStas Monte Carlo computer simulations were conducted with the results presented in figure 11. These pictures show an effectively identical $AD_S$ (*ie* the same $A,B,C$ chosen here to be suitable for a high resolution PD) produced by widely different PS configurations. The simulated transmission over a 10 mm wide sample is plotted as a greyscale map of $\tau(\kappa_x/\kappa, \Delta\kappa/\kappa)$. The solid and dotted line elliptical contours represent the calculated $\tau=1/2$ and $\tau=0.0625$ contours for Gaussian elements. The vertical dotted lines represent the limits imposed by $\alpha_1$ while the pairs of sloping lines represent the limits imposed by $\alpha_0$ and $\eta_M$. All these simulations used $\lambda=1.5$ Å, $L_1=2.0$ m and a 15 cm wide source. The reactor face was modelled at 4 m from the source with the monochromator some distance further.

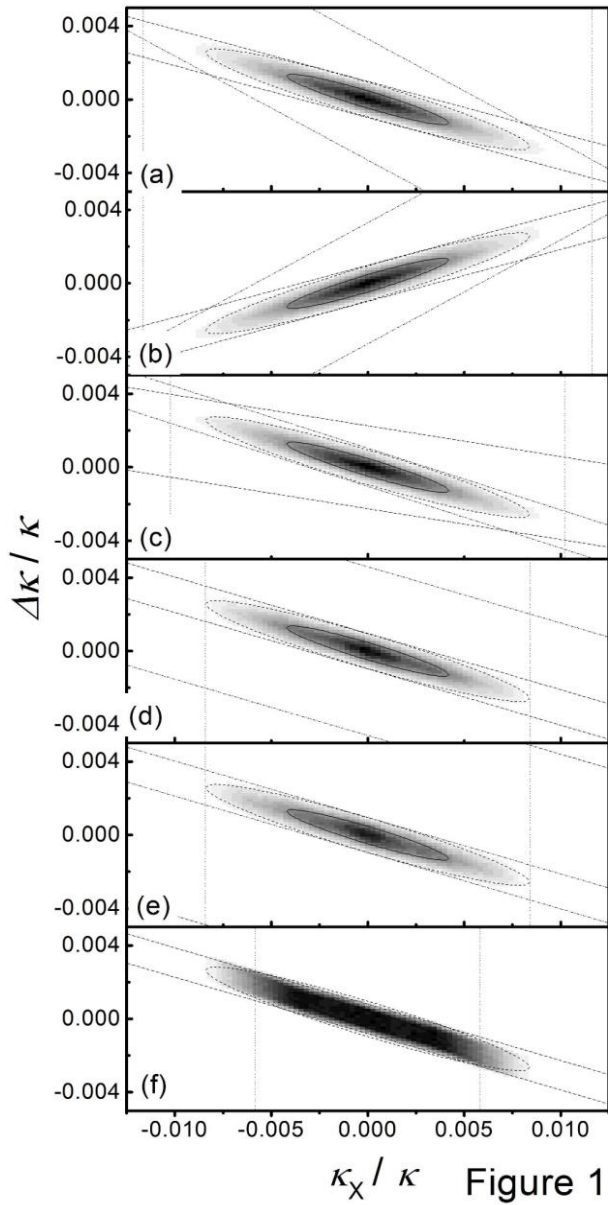

Figure 11

Figures 11a-c show configurations using Soller collimators and a flat mosaic monochromator. Figures 11a and 11b use values $\{\alpha_0,\eta_M,\alpha_1\} = \{12',20',40'\}$ with $\theta_M= 60.51°$ and -60.51° respectively corresponding to a germanium (Ge533) monochromator. Note that reversing the sign of $\theta_M$ reverses the AD slope. This simply means that if $\theta_M$ is negative, decreasing $\theta_M$ increases the magnitude of $\theta_M$ and hence decreases the magnitude of $\kappa$. Figure 11c uses $\{\alpha_0,\eta_M,\alpha_1,\theta_M\} = \{46.2',10.7',35.1',71.44°\}$, *ie* Ge711, to produce an identical beam.

Figures 11d and 11e show configurations using a Radial Soller $\alpha_0$, a Soller collimator $\alpha_1$ and a mosaic HFM with $\{\alpha_0,\eta_M,\alpha_1,\theta_M,L_0,R_{MH}\}= \{8.43',20',28.9',51.75°,8.95m,4.163m\}$ (Ge531) and $\{6.3',20',28.9',43.6°,4.8m,4.1m\}$ (Ge 511) respectively.

Figure 11f uses open beamtubes defined by slits at a virtual source and before the monochromator with a mosaic HFM. This arrangement gives results similar to that for the Radial Soller-flat monochromator-Soller arrangement discussed above but delivers rectangular profile $\tau(\gamma_0)$ and $\tau(\gamma_1)$. The monochromator modelled used a Gaussian mosaic which means that $AD_S$ does not have a fully rectangular transmission profile. Here $\{\eta_M,\theta_M,R_{MH}\} = \{20',51.75°,4.163m\}$, the virtual source is 0.031 m wide at 8.95m from the monochromator and the slit before the monochromator is 0.0235 m wide. These values give element angular divergence HW a factor of √2 smaller than the element FWHMs used for fig. 11d. This choice is to ensure that the rectangular profiles generated by the slits have the same angular variance as do the triangular profiles in fig 11d. All curved monochromators were 15 cm wide and composed of 15 segments. The simulated flux at the sample was the same for the examples in fig 11a-e within 15% (as would be expected for equivalent $AD_S$) but 135% higher for fig 11f. All collimators and monochromators are assumed to have the same 100% peak transmission. The angular and average wavevector variances were extracted from the figure 11 data and the deduced standard deviations are tabulated in table 1. The largest variation is 7.6% showing that $AD_S$ is the same in effect for these very different PS configurations.

It is possible to use even smaller values for $\theta_M$ to match $AD_S$ here and some calculated examples for a Radial Soller – mosaic HFM – Soller arrangement are

- $\{\alpha_0,\eta_M,\alpha_1,\theta_M,L_0,R_{MH}\} = \{3.20',20',28.9',26.1°,2.86\ m,5.34\ m\}$ (Ge311)

- $\{\alpha_0, \eta_M, \alpha_1, \theta_M, L_0, R_{MH}\} = \{1.49', 20', 28.9', 12.9°, 2.33\text{ m}, 9.63\text{m}\}$ Pyrolytic Graphite PG002
- $\{\alpha_0, \eta_M, \alpha_1, \theta_M, L_0, R_{MH}\} = \{1.49', 20', 28.9', -12.9°, 1.75\text{ m}, 8.35\text{m}\}$ PG002

| Figure | $\sigma(\kappa_x/\kappa)$ | $\sigma_{Av}(\Delta\kappa/\kappa)$ |
|---|---|---|
| 11a | 0.0155 | 0.00166 |
| 11b | 0.0154 | 0.00166 |
| 11c | 0.0148 | 0.00162 |
| 11d | 0.0145 | 0.00168 |
| 11e | 0.0145 | 0.00178 |
| 11f | 0.0157 | 0.00174 |

Table 1: Standard deviations for wavevector spreads in figure 11.

This last example combines a negative $\theta_M$ with $L_1 > L_0$ to produce an unchanged $AD_S$ slope. Simulations of these configurations showed that $AD_S$ matched those in figure 11 closely if the sample width was 1 mm but a 10 mm sample width showed a larger $\Delta\kappa$ width and $AD_S$ edges showing a noticeable curvature as illustrated in figure 12. These examples use small $\alpha_0$ and large $\eta_M$. Since the important quantity here is $(4\alpha_0^{-2}+\eta_M^{-2})$ it should be possible to equivalently use large $\alpha_0$ and small $\eta_M$ but simulations showed that this approach led to even larger $AD_S$ distortions. Figure 11 does demonstrate that an HFM can be used to greatly increase the $\theta_M$ range one can use effectively which may be very convenient for a number of practical reasons. Figure 12 shows that there is a limit to the theory where a check of spatial uniformity in the beam shows problems. Admittedly, the figure 11 $AD_S$ represents a rather extreme and very specialised beam type as required for a high resolution powder diffractometer to be used with a large range of sample scattering angle. The changes from very large $\theta_M$ in figure 11a to the very small $\theta_M$ in figure 12b and reversed small $\theta_M$ in figure 12c is also quite extreme. The origin of the aberrations in figure 12 is not clear. Reference [12] shows that decreasing $L_0$ in an open geometry PS increases the contributions from source width to the spatial and angular width of the sample beam spot size. Decreasing $L_0$ increases the contribution to the $\Delta\kappa$ width from the source width while the contribution from monochromator width passes through a minimum at $L_0=L_1$.

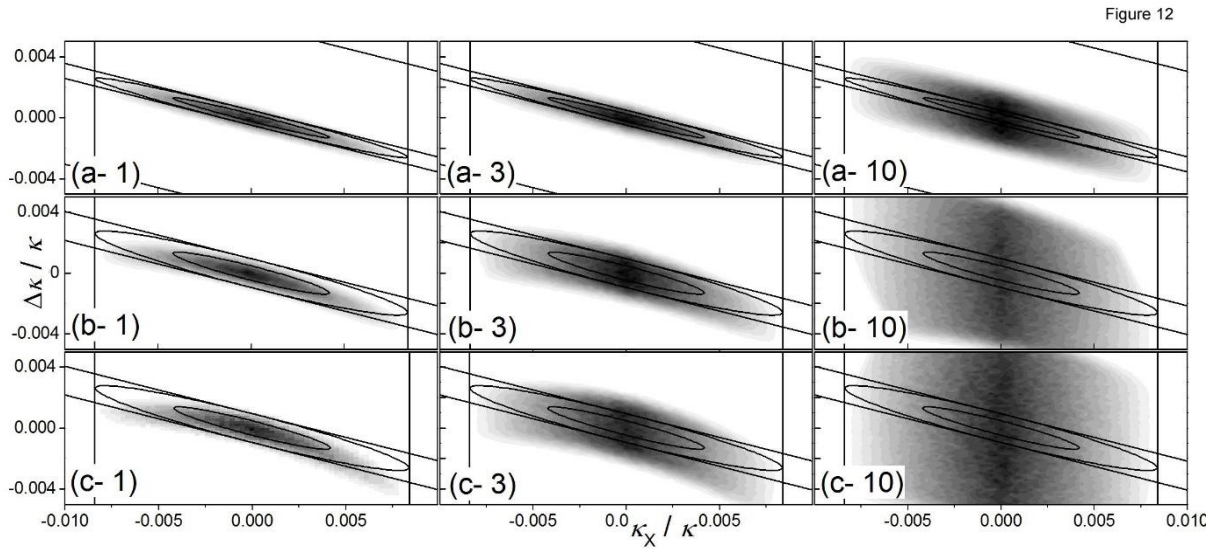

Figure 12

The monochromators modelled in these simulations use segments on a flat base oriented to give angular displacements following a cylindrical curvature, so the monochromator's curvature is not even truly cylindrical. It may be that a truly cylindrical curvature or perhaps some other curvature would reduce the aberrations seen here but such a study is beyond the scope of this work.

The exact equivalence of the PS AD for different choices of instrument variables only applies to beam elements with exactly Gaussian transmission functions. For triangular and rectangular distributions the "equivalence" is only approximate but the figure 11 simulations show that it is "closely approximate"; perhaps better phrased as "usefully approximate".

The fact that different choices of beam elements can deliver the same beam characteristics at the sample has academic interest but should also have practical use in overcoming the limitations inherent in available components. It is simple to continuously vary a slit width, vary monochromator Bragg angle or exchange collimators between discrete values. It is difficult to make high transmission Soller collimators with small divergence but simple to make a narrow slit (with high transmission). It is usually difficult to make high peak reflectivity monochromators with large mosaic and a given monochromator crystal may only be available with some particular mosaic. It is possible to interchange monochromators to change $\eta_M$ or $d_M$ or to adjust the relationship between $\lambda$ and $\theta_M$. It has been shown here that wave-vector spread can be adjusted by an $\alpha_0$ slit width without changing $\eta_M$. Using a curved HFM with smaller $\eta_M$ at a smaller $\theta_M$ can produce the same beam effect as a flat monochromator with larger $\eta_M$ at larger $\theta_M$. Using slits to produce rectangular transmission profiles which increase intensity for a given resolution is simpler, cheaper and more controllable than using guides or reflecting Soller collimators. Background considerations may affect the choices but these are difficult to predict and very difficult to simulate accurately.

Most importantly from the motive for this work, a given $AD_S$ can be produced by different choices of primary spectrometer elements. Therefore, attempting to optimise an instrument design using the beam elements as variables must be hindered by the significant parameter covariance. This can be avoided by using a set of variables which describe the beam character such as $(u, \psi_{AD}, \alpha_{In})$. No doubt other variable choices could be made if desired.

## 7. Adjusting the Primary Spectrometer to achieve a desired beam character

Section 6 showed that many different PS configurations can deliver equivalent beams. This section shows that a single PS design can deliver widely different beams as illustrated by $AD_S$. There are many ways of dealing with the choices available to deliver a given beam character but this section only discusses two. The first uses a PS with open beamtube $\alpha_0$ and $\alpha_1$ with a mosaic HFM and with $AD_{\alpha 0}$ and $AD_{Mono}$ matched in slope. The second examines a PS using a HFM of small mosaic on a guide of rather large $m\theta_C\lambda$ giving relatively large beam angular divergence.

Figure 13 shows McStas simulations of a PS using open beamtubes and a mosaic HFM as an independent confirmation of the calculations presented above. Figure 13a acts as a reference point and shows $AD_S$ over a 10 mm width for a PS using Soller collimators and a flat mosaic monochromator with $(\alpha_0, \eta_M, \alpha_1, \theta_M, L_0, L_1, R_{MH}) = (20', 24', 30', 20.6°, \infty, 2.2\ m, \infty)$ at $\lambda=2.36$Å (PG002 monochromator).

Figure 13b shows $AD_S$ for a PS with $(\alpha_0, \eta_M, \alpha_1, \theta_M, L_0, L_1, R_{MH}) = (15', 24', 30', 20.6°, 10m, 2m, 9.47m)$. There is much to note in fig 13b. Firstly, the open beamtubes lead to rectangular profiles. This is limited slightly by the Gaussian monochromator mosaic modelled. $\alpha_0$ is achieved using $W_V = \pm 4.36$ cm and $L_0=10$ m setting a rectangular profile of HW $15' = 21'/\sqrt{2}$. $\alpha_1$ is set by a slit of width 3.8 cm just before the monochromator with $L_1=2.0$ m giving a rectangular HW 30'. The rectangular profiles result in an $AD_S$ slope which differs visibly from that in fig 13a as would correspond to a larger $\theta_M$. Note that the value of $\alpha_0$ limits the $\Delta\kappa/\kappa$ variation but that the 24' Gaussian monochromator mosaic causes a rounding of the top of $AD_S$.

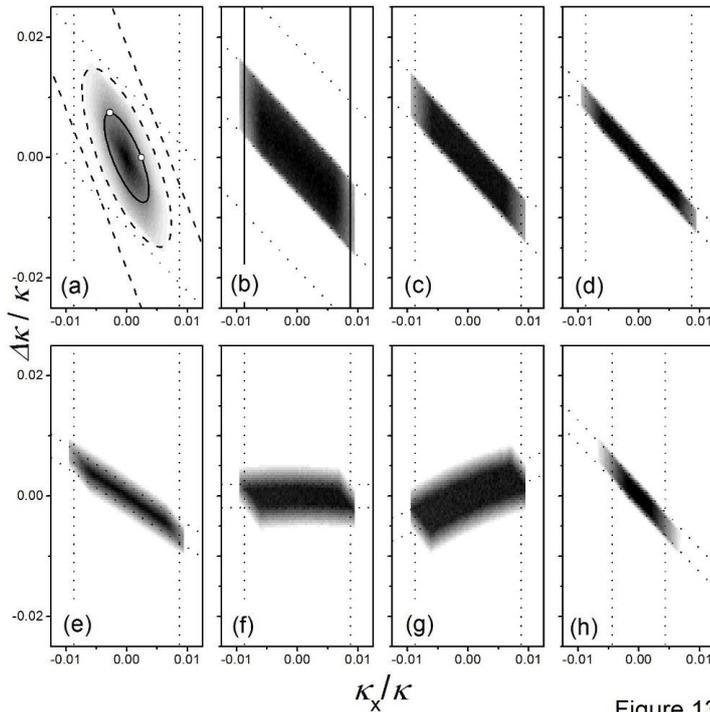

Figure 13

Fig 13c shows the effect of increasing the monochromator mosaic to 50' (flattening the peak transmission of $AD_S$ and increasing its average) while reducing $\alpha_0$ to 10' rectangular HW. Fig 13d shows the effect of reducing $\alpha_0$ to 5'. This clearly demonstrates that if a large enough monochromator mosaic is chosen, the $\Delta\kappa/\kappa$ width at the vertical axis of $AD_S$ is effectively fully controlled by $\alpha_0$, i.e. by the virtual source slit width. Note that this is independent of the slope of $AD_S$.

Figs 13e, f & g show that varying $L_0$ (to 4.0, 2.0 and 1.5 m) while keeping $\alpha_0 = 5$' alters the slope of $AD_S$. Note that at each value of $L_0$, $W_V$ must be adjusted to keep $\alpha_0 = W_V/L_0$ constant at 5' and $R_{MH}$ is adjusted according to eq. 12 to maintain conventional full focussing. Note that the slope of $AD_S$ can be varied and even reversed while keeping $\theta_M$ (and hence $\lambda$) fixed. Varying $\theta_M$ would alter the $AD_S$ slope and $AD_S$ $\Delta\kappa/\kappa$ width (by a $\cot\theta_M$ factor) as well as changing the wavelength. The theory says that in this configuration with a large $\eta_M$, $\alpha_0$ (here controlled by the slit width $W_V$) should control the $\Delta\kappa/\kappa$ width. Figure 13f shows a significant $\Delta\kappa/\kappa$ extra broadening for the 1 cm wide sample. This is not the case for smaller sample widths and thus, for "extreme" focussing conditions where $L_0$ becomes small the sample width may have to be reduced to reduce the wave-vector spread $\Delta\kappa$ (corresponding to energy width in the scans). There may be circumstances where this spread is acceptable.

Fig 13h shows that $\alpha_1$ (controlled using a slit before the monochromator) controls the overall $\gamma_1$ width (and $\Delta\kappa_x/\kappa$ width) of $AD_S$. Here, $a_1 = 15$'. The intensities observed over a 1 cm width sample for figures 13a-h are in the ratio 1 : 4.06 : 1.55 : 0.84 : 0.48 : 0.36 : 0.32 : 0.45. Notice that for the plots showing a change in the ratio $L_0/L_1$ with a corresponding change in the monochromator curvature (figures 13d, e, f & g), a larger ratio $L_0/L_1$ results in larger sample flux.

Figure 13 shows that the behaviour of $AD_S$ with rectangular transmission profiles closely follows that expected from the relations in section 4. Adjusting the virtual source width changes the $AD_S$ width measured parallel to the $\Delta\kappa$ axis with no other effect. Adjusting $\alpha_1$ using $W_M$ varies the $AD_S$ calliper width parallel to the $\kappa_x$ axis. Adjusting $L_0$ (while simultaneously varying $W_V$ to maintain $\alpha_0$ and varying $R_{MH}$ to maintain source to sample focussing) changes the slope of $AD_S$. At small ratios of $L_0/L_1$, significant aberrations appear resulting in a $\Delta\kappa/\kappa$ broadening. This is not evident for a 1 mm wide sample beam but is apparent for the 10 mm wide beam shown here. Tests showed that the $AD_S$ distortions observed here appear to be smaller at larger values of $\theta_M$.

For a primary spectrometer situated on a guide source it is effectively impossible to vary $L_0$ so one must accept the $AD_{\alpha 0}$ slope supplied by the guide. If the instrument is sited at the end of the guide, it may be possible to position a virtual source at the guide end and site the monochromator at some distance to exploit the flexibility offered by this arrangement. Such an arrangement was found to be useful in a purely numerical optimisation discussed [9]. For instruments on non-end guide positions $\alpha_0$ within the guide is set by the choice of $\lambda$, usually to a rather small value. In some cases, it may be possible to use interchangeable collimators between guide and monochromator to reduce the $\Delta\kappa/\kappa$ width of $AD_S$. By using a monochromator of fairly small mosaic one can control the $AD_S$ slope by varying $R_{MH}$. A slit just after the monochromator can be used to control $\alpha_1$ and thus the $\gamma_1$ width of $AD_S$. Figure 14 illustrates the effect of varying the curvature of a HFM on a long guide. The $AD_S$ simulated and illustrated here uses a 3 cm wide $m=3$ guide at 2.36Å giving $\alpha_0 = \pm\, m\theta_c\lambda = \pm 0.71°$ or 42'. A Ge111 monochromator is modelled with $\eta_M=12'$ and $\theta_M=21.21°$. With $L_1 = 1.72$ m, $\alpha_1 = \pm 27'$ with a rectangular profile. The monochromator has variable curvature with values in figure 14 of (a) Flat; (b) $R_{MH} = 9.51$m (eq. 12 focussing) (c) $R_{MH} = 4.75$m and (d) $R_{MH} = 3.17$m. This figure illustrates very clearly that $AD_S$ results from the superposition of $AD_{\alpha 0}$, $AD_{Mono}$ and $AD_{\alpha 1}$. The relative intensities over a 1 cm wide sample are 1 : 1.13 : 1.08 : 0.77 showing that the main effect of a HFM is to alter the $AD_{Mono}$ slope with any intensity variation resulting from a changing overlap with $AD_{\alpha 0}$. The $AD_S$ reversal shown in fig 14d may be useful. For instruments on a guide using small $\theta_M$, it is difficult to use the "zigzag" focussing arrangement for monochromator and sample scattering as the secondary spectrometer would collide with the guide shielding. Reversing the $AD_S$ slope reverses the sample scattering sense needed to focus scans. It is likely that the increased intensity from focussing a scan would outweigh any loss in sample position flux from this approach.

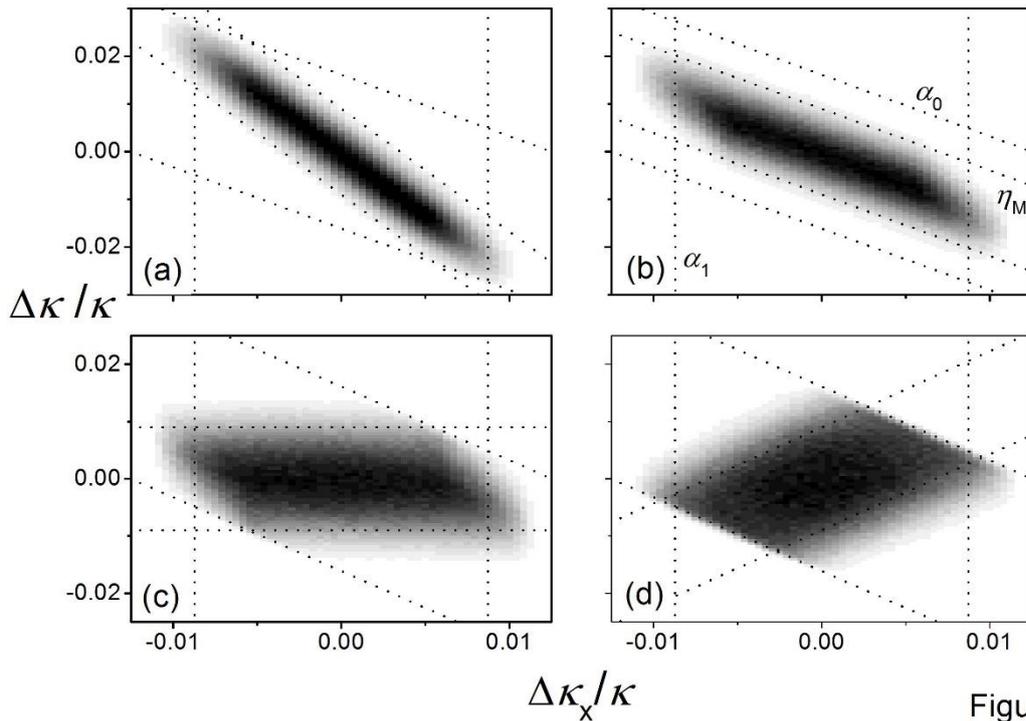

Figure 14

If working on a beamtube rather than a guide, one way to reverse the $AD_S$ slope is to fix $\alpha_0$, choose a small $\eta_M$ and vary $R_{MH}$ to rotate $AD_{Mono}$. One could set $L_0=L_1$ to set the $AD_{\alpha 0}$ axis parallel to the $\gamma_1$ axis. Another way to achieve the same end would be to use a large $\eta_M$ with $R_{MH}=L_1/\sin\theta_M$ to set the $AD_{Mono}$ axis parallel to the $\gamma_1$ axis and then use a small $\alpha_0$ and vary $L_0$ to rotate $AD_{\alpha 0}$.

## 8. A Modified Primary Spectrometer Design

There are a number of technical difficulties in designing conventional primary spectrometers.

- The large beamtubes needed to supply large divergence and high intensity also almost inevitably give high background.

- Increasing beamtube size leads to flux depression at reactor sources.

- If using Soller collimators to deliver small angular divergences the peak transmission declines.

- High resolution measurements usually require relatively large values of $\theta_M$.

- Large mosaic monochromators tend to have low reflectivity due to incoherent scattering and absorption within the monochromator.

- Air scattering in long flight paths significantly reduces beam flux and increases background and shielding requirements.

- It is difficult to calculate, model or simulate the background produced in a PS.

In general, it is desirable that a given PS is capable of delivering a range of resolution characteristics. An ideal PS would be fully and simply controllable to deliver the maximum possible transmission with a low background for a range of resolution options and have simply modelled resolution characteristics. Rectangular $\tau(\gamma)$ distributions deliver higher total transmission for a given angular variance than do Gaussian or triangular profiles. Open beam tubes deliver locally rectangular $\tau(\gamma)$. Such profiles can be expected to deliver measured scan peaks with non-Gaussian angular distributions. While that may be aesthetically displeasing, since all but the simplest data is now routinely analysed by computer modelling, this is not a serious disadvantage as long as the profiles are accurately calculable or can at least be modelled. Traditionally Soller collimators and monochromator crystals were exchanged to vary instrument resolution. It would be better to be able to simply and remotely vary some PS parameters continuously to adjust the beam character in a known and well understood way.

All the adjustments described here could be achieved using the PS design illustrated in figure 15 where a source is followed by a heavy slit (virtual source) with both variable slit width, $\pm W_V$, and variable position between the source and monochromator. The distance from virtual source to monochromator, $L_0$, controls the slope of $AD_{\alpha 0}$, $\left[-\frac{1}{2}\left(1-\frac{L_1}{L_0}\right)cot\theta_M\right]$, and $W_V$ controls the value of $\alpha_0$ ($\approx \pm W_V/L_0$). In practice, the heavy $W_V$ slit could be mounted on rails in an open beamtube between the source and the monochromator although the very heavy fixed shielding needed near the source may limit the range over which $L_0$ could be varied. A second heavy slit just before (or just after) the monochromator controls $W_M$ and hence $\alpha_1$ ($\approx \pm W_M \sin\theta_M/L_1$). One challenge is that if $L_0$ is made very small, a large source width is needed to fully illuminate sufficient monochromator width to permit large values for $\alpha_1$. The $W_V$ and $W_M$ slits could both be made very heavy and since both are inside the heavy monochromator and beamtube shielding this should effectively reduce background from fast neutrons and gamma rays from the source. The path between the monochromator and sample is also an open beamtube where the beam angular width is controlled by the variable slit before the monochromator. It would obviously be advisable to include shielding or reduce air scattering where possible in this region. Since mosaic cannot be altered in situ, ideally a rather large mosaic could be chosen and $\alpha_0$ ($W_V$) used to control $\Delta\kappa/\kappa$. For many monochromator materials, increasing $\eta_M$ reduces peak reflectivity but note that the important quantity here is $\eta_M \cot\theta_M$ so at small $\theta_M$, great versatility may be possible even with a relatively small $\eta_M$. Adjusting $R_{MH}$ and $L_0$ in concert matches the slope of $AD_{Mono}$ $\left[-\left(1 - \frac{L_1}{R_{MH}\sin\theta_M}\right)\cot\theta_M\right]$ to that of $AD_{\alpha 0}$ to maximise transmission (although other choices could be made). Since the beamtubes and slits impose little restriction, much of the path length could be evacuated or filled with He gas to reduce air scattering.

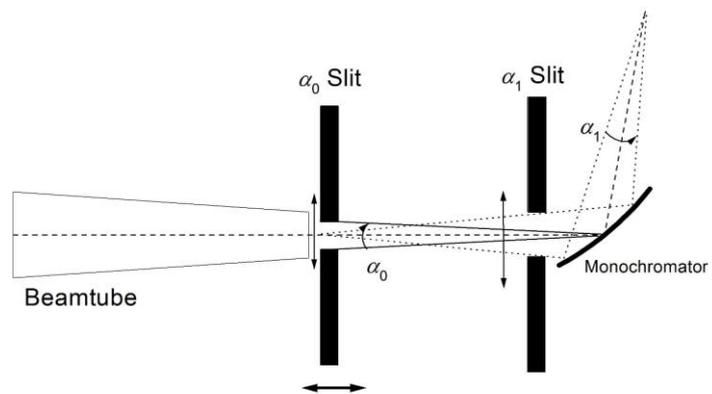

Figure 15

The slope of $AD_S$ is controlled by the matched slopes of $AD_{\alpha 0}$ and $AD_{Mono}$ set by $R_{MH}$ and the ratio $L_1/L_0$. The $\gamma_1$ width of $AD_S$ is $\pm W_M \sin\theta_M / L_1$ controlled by the width $W_M$. The $\Delta\kappa/\kappa$ width of $AD_S$ at $\gamma_1=0$ is controlled by the width $W_V/L_0$ (combined with $\eta_M$) and if $\eta_M$ is large is $\approx \pm \frac{1}{2} W_V \cot\theta_M / L_0$. This design offers great flexibility in choosing $AD_S$ and the rectangular element profiles should maximise transmission.

The key point illustrated here is that an understanding of the form of $AD_S$ and its origin in PS elements allows the design of a PS with simply adjusted instrument parameters to achieve a wide range of beam character at the sample. To best set up scans – to optimise instruments or measurements – now only requires knowledge of the beam character needed.

## 9. Discussion and Conclusion

This work presents equations to find the primary spectrometer variables needed to deliver a sample position beam of some desired character. There are usually many such choices of primary spectrometer beam elements possible. That multiplicity means that there is significant parameter covariance in any attempt to optimise a neutron scattering instrument if using the instrument parameters as the variables. Choosing suitable parameters describing the beam removes this covariance from the problem.

The approach adopted to this was to extend the 2D Acceptance Diagram, plots of $\tau(\kappa_x, \Delta\kappa)$, to include a range of collimation options and HFMs. This approach sheds light on the effect of individual PS elements and one result of the visual presentation is the flexible new PS design with maximised transmission described in section 8. The total sample position AD can be viewed as the product or superposition of three ADs each associated with a single PS element and curving the monochromator shears $AD_{Mono}$ parallel to the $\Delta\kappa/\kappa$ axis which increases the overlap between $AD_{\alpha 0}$ and $AD_{Mono}$ and thus transmission. The important assumption adopted that the beam at the sample centre is representative of the whole beam appears to be valid over a wide range of parameter choices but may break down in some cases. Since a non uniform sample beam may affect the measurements, identifying such cases acts as a warning.

This visual AD view of beam character suggests a number of potential benefits and clarifies some known effects.

- While simple and radial Soller collimators deliver beams with triangular local transmission profiles, $\tau(\gamma)$, guide tubes and beamtube collimators produce rectangular transmission profiles and can be used to increase count rates for given scan angular widths.

- Using an HFM at small $\theta_M$ can reduce the need for monochromators with large mosaic.

- Using an HFM on 4 circle SXD's should allow full wave-vector focussing at peaks with widely different $d_S$ even at fixed $\theta_M$.

- CW PDs often need good resolution over a large scattering angle, $\theta_S$, range. Conventionally, this is achieved by using a rather large monochromator Bragg angle where $\cot\theta_M$ is small giving an $AD_S$ with a small slope (large $\psi_{AD}$) and a small $\Delta\kappa/\kappa$ width. An HFM can give an equivalent effect at smaller $\theta_M$ and this may be mechanically convenient.

- A single CW PD using the novel PS described here should be operable in high and low resolution modes by simply adjusting two slits to vary $W_V$ and $W_M$.

The AD pictures suggest ways to visualise and perhaps better exploit a number of known beneficial approaches to instrument design and use:

- HFMs can be used to tailor beams to have a very large overall $\Delta\kappa/\kappa$ with an $AD_S$ slope chosen to deliver very high intensity over a small $\theta_S$ range at some desired $\theta_S$ (as is required, for example, on strain scanning instruments).

- TAS analysers are just primary spectrometers in reverse and all of the mathematics presented here should apply equally to a TAS analyser with a reversal of neutron path (*ie* AD slope).

- It should be possible to maintain the shape of $AD_S$ as $\theta_M$ is changed to vary the incident wavevector in a scan (on TAS for example) requiring a change in $\theta_M$) meaning that scans could be conducted with unvarying resolution. While this is impractical with a beam defined by collimators, it should be possible for a primary spectrometer (and / or secondary spectrometer if necessary) using open beam tubes with continuously variable slit widths at the source and monochromator and if the ratio $L_1/L_0$ can be varied.

- For TAS designed to operate with "monochromatic" focussing, varying the virtual source width directly controls the energy resolution for small samples (with a variation in intensity of course).

- Varying the $AD_S$ slope in a controlled way should make it possible to focus any TAS scan of acoustic excitations.

- HFM curvature can be used to reverse the $AD_S$ slope thus changing the focussed instrument side in a scan as would be useful for TAS on guide tube sources where the usually favoured "W" configuration is inconvenient at small $\theta_M$.

- It may be possible to incorporate this AD formalism in computer simulation programs for instruments which should be much faster than existing MC programs and facilitate the use of such simulation programs as the kernel of an optimisation routine.

- Nowhere in this work is it essential that the radiation considered be neutrons so the formalism should work for X-Ray or electron scattering instruments if the instrument resolution is dominated by the beam elements.

Others may find further applications for this formalism. These AD pictures are simple to imagine and draw and can be used to describe the behaviour of conventional primary spectrometers: the rather tedious mathematical derivations are only needed to draw the pictures accurately. These pictures clearly illustrate that an HFM's primary effect is to alter the correlation between angle and wave-vector in the beam at the sample. The pictures here deal only with the in-plane 2D AD and while this describes the most complex part of PS effects, any full description must include a consideration of vertical divergence effects and possibly also spatial effects. A 3D $\kappa$-space volume can be constructed by including any beam vertical divergence which is largely decoupled from the in-plane effects. The full instrument resolution must also consider sample scattering and the secondary spectrometer resolution. Being able to describe $AD_S$ and choose PS elements to deliver a desired $AD_S$ means that maximising instrument performance only requires knowledge of the $AD_S$ needed. While the PS described here is common to SXD's, CW PDs and TAS, the $AD_S$ required depends on the measurement and this rather important question of which choice of $AD_S$ is best will be addressed elsewhere. Proper instrument design requires more than just a consideration of beam intensity at the sample. A proper understanding of beam character at the sample and its production also shows how to maximise intensity. It is hoped that the formalism presented in this work will provide a useful tool for the better design of neutron scattering instruments.

**Acknowledgement**: LDC thanks Dr. Manh Duc Le for useful discussions and Peter Willendrup who collaborated on some early MC simulations of some of the effects discussed here.

**Figure Captions**

**Figure 1**     Schematic diagram of the primary spectrometer discussed.

**Figure 2**     This diagram illustrates the acceptance diagram view of the sample beam. The pictures describe the transmission in a 2D wave-vector space with *x*-coordinate the transverse wave-vector component and *y*-coordinate the variation from nominal longitudinal wave-vector component. The total transmission $\tau(\kappa_x, \Delta\kappa)$ shown in figure 2d is the product of $AD_{\alpha 0}$ (fig. 2a), $AD_{Mono}$ (fig. 2b) and $AD_{\alpha 1}$ (fig. 2c) which describe respectively the transmission effects of the pre-monochromator collimator $\alpha_0$, the monochromator and the pre-sample collimation, $\alpha_1$.

**Figure 3** Phase space diagrams showing the transmission of various collimator types as a function of position and angle – the angle is proportional to the transverse wave-vector (or momentum) component. $\tau(\gamma,x)$ for (a) a guide (b) a Soller collimator (c) a diverging radial Soller collimator (d) a converging radial Soller collimator and (e) an open beam tube or separated slit pair.

**Figure 4** Transmission for an open beam tube (a separated pair of slits) here of equal width.

(a) $\tau(\gamma,x)$ immediately after the second slit ;

(b) $\tau(\gamma)$ after the second slit ;

(c) $\tau(\gamma,x)$ some distance behind the second slit.

**Figure 5** DuMond Diagrams – plots of $\tau(\gamma, \lambda)$

    (a) $\alpha_0 = \eta_M = 0$                  (b) $\alpha_0 = 0$ ; finite $\eta_M$ – dotted line shows $\eta_M = \pi$

    (c) Finite $\alpha_0$ ; $\eta_M = \pi$            (d) $\alpha_0 = \pi$ ; Finite $\eta_M$

    (e) Finite $\alpha_0$ ; Finite $\eta_M$        (f) $\alpha_1$ collimator limits beam angular width at sample.

    (g) $d_M$ gradient monochromator (h) $\alpha_1$ collimator plus $d_M$ gradient monochromator

**Figure 6** Acceptance Diagrams – plots of $\tau(\gamma, \kappa)$

(a) $\tau(\gamma, \kappa)$ for finite $\alpha_0$ ; mosaic monochromator with finite $\eta_M$
(b) $\tau(\gamma, \kappa)$ for finite $\alpha_0$ ; gradient monochromator with finite $\mu_M$

**Figure 7** The broadening effects of

(a) $\alpha_0$ on $AD_{\alpha 0}$. $AD_{\alpha 0}$ centreline is $\Delta\kappa/\kappa = -\frac{1}{2}\gamma\cot\theta_M$ found by setting $\alpha_0 = 0$ and $\eta_M = \infty$. For flat monochromators (illustrated here) broadening has extent $\pm\frac{1}{2}\alpha_0$ (// $\kappa$ axis) along $\Delta\kappa/\kappa = -\gamma\cot\theta_M$.
(b) $\eta_M$ on $AD_{Mono}$. $AD_{Mono}$ centreline is $\Delta\kappa/\kappa = -\gamma\cot\theta_M$ found by setting $\eta_M = 0$ and $\alpha_0 = \infty$. For flat monochromators as here broadening is $\pm\eta_M$ (// $\kappa$ axis) along $\Delta\kappa/\kappa = -\frac{1}{2}\gamma\cot\theta_M$.

**Figure 8** McStas simulations showing $AD_S$ $AD_S(\Delta\kappa_x/\kappa, \Delta\kappa_z/\kappa)$ illustrating the broadening effect of

(a) $\alpha_0$ on $AD_{Mono}$ where $\alpha_0 = 3$' with $\eta_M = 20$'
(b) $\eta_M$ on $AD_{\alpha 0}$ where $\eta_M = 3$' with $\alpha_0 = 30$'

**Figure 9** Illustration of a primary spectrometer $AD_S$ arising from Gaussian profile collimators and mosaic showing $u$, $\psi$ & $\alpha_{In}$ with their extent indicated by dotted lines. The two ellipse contours represent the 50% and 6.25% transmission contours.

**Figure 10**

(a) Values of $(\alpha_0, \eta_M, \alpha_1)$ vs $\theta_M$ for given $(A, B, C)$ in a Soller-flat mosaic-Soller PS. The values shown here deliver a beam identical to that for $(\alpha_0, \eta_M, \alpha_1, \theta_M) = (12', 20', 40', 60°)$
(b) Values of $(\alpha_0, \eta_M, \alpha_1)$ vs $\theta_M$ for given $(A, B, C)$ in a Soller-flat mosaic-Soller PS. The values shown here deliver a beam identical to that for $(\alpha_0, \eta_M, \alpha_1, \theta_M) = (10', 20', 40', 60°)$. Note the split in the allowed range of $\theta_M$.

(c) Values of $\alpha_1$, $L_1/L_0$ and $(4\alpha_0^{-2}+\eta_M^{-2})$ vs $\theta_M$ for given (A, B, C) in a Radial Soller-curved mosaic-Soller PS. Here the beam delivered matches that for fig 9a.

**Figure 11** McStas simulations of the sample position beam $AD_S$ for different sets of instrument variables showing that a range of primary spectrometer arrangements and variable sets can deliver effectively identical beam characteristics. Figures 11a-c use a conventional PS with Soller collimators and a flat mosaic monochromator. Figures 11d-e use a PS with a Radial Soller - curved mosaic monochromator and Soller collimator. Figure 11f uses a PS with open beamtubes and a curved mosaic monochromator. All examples deliver a $\lambda=1.5$Å beam and use $L_1=2.0$ m.

(a) $(\alpha_0, \eta_M, \alpha_1, \theta_M) = (12', 20', 40', 60.5°)$
(b) $(\alpha_0, \eta_M, \alpha_1, \theta_M) = (12', 20', 40', -60.5°)$
(c) $(\alpha_0, \eta_M, \alpha_1, \theta_M) = (46.2', 10.7', 35.1', 71.4°)$
(d) $(\alpha_0, \eta_M, \alpha_1, \theta_M) = (8.43', 20', 28.9', 51.8°)$ and $(L_0, L_1, R_{MH}) = (8.95, 2.0, 4.16)$
(e) $(\alpha_0, \eta_M, \alpha_1, \theta_M) = (6.27', 20', 28.9', 43.6°)$ and $(L_0, L_1, R_{MH}) = (4.8, 2.0, 4.09)$
(f) $(\eta_M, \theta_M) = (20', 51.8°)$ and $(L_0, L_1, R_{MH}) = (8.95, 2.0, 4.16)$ using a 3.1 cm wide virtual source and a 2.4 cm wide slit just before the monochromator.

**Figure 12** McStas simulations of the sample position beam $AD_S$ using a PS with a Radial Soller - curved mosaic monochromator and Soller collimator with $\lambda=1.5$Å beam and $L_1=2.0$ m. These examples show the effect of sample size (from left to right - 1mm, 3 mm and 10 mm) with small $\theta_M$.

(a) $(\alpha_0, \eta_M, \alpha_1, \theta_M) = (3.2', 20', 28.9', 26.12°)$ and $(L_0, L_1, R_{MH}) = (2.86, 2.0, 5.35)$ (Ge311)
(b) $(\alpha_0, \eta_M, \alpha_1, \theta_M) = (1.5', 20', 28.9', 12.92°)$ and $(L_0, L_1, R_{MH}) = (2.33, 2.0, 9.63)$ (PG002)
(c) $(\alpha_0, \eta_M, \alpha_1, \theta_M) = (1.5', 20', 28.9', -12.92°)$ and $(L_0, L_1, R_{MH}) = (1.75, 2.0, -8.35)$ (PG002)

**Figure 13** McStas simulations of $AD_S$ for a PS with aligned $AD_{\alpha 0}$ & $AD_{Mono}$ in various configurations. The superimposed lines show the expected full width contours for $AD_{\alpha 0}$, $AD_{Mono}$ and $AD_{\alpha 1}$. Figure 13a shows $AD_S$ for a conventional PS using Soller collimators and flat monochromator while figures 13b-h show $AD_S$ for a HFM virtual source PS using beamtube collimation.

(a) $(\alpha_0, \eta_M, \alpha_1, \theta_M) = (30', 24', 30', 20.6°)$.
(b) $(\alpha_0, \eta_M, \alpha_1, \theta_M, L_0, L_1, R_{MH}) = (15', 24', 30', 20.6°, 10m, 2m, 9.47m)$
(c) $(\alpha_0, \eta_M, \alpha_1, \theta_M, L_0, L_1, R_{MH}) = (10', 50', 30', 20.6°, 10m, 2m, 9.47m)$.
(d) $(\alpha_0, \eta_M, \alpha_1, \theta_M, L_0, L_1, R_{MH}) = (5', 50', 30', 20.6°, 10m, 2m, 9.47m)$.
(e) $(\alpha_0, \eta_M, \alpha_1, \theta_M, L_0, L_1, R_{MH}) = (5', 50', 30', 20.6°, 4m, 2m, 7.58m)$.
(f) $(\alpha_0, \eta_M, \alpha_1, \theta_M, L_0, L_1, R_{MH}) = (5', 50', 30', 20.6°, 2m, 2m, 5.68m)$.
(g) $(\alpha_0, \eta_M, \alpha_1, \theta_M, L_0, L_1, R_{MH}) = (5', 50', 30', 20.6°, 1.5m, 2m, 4.87m)$.
(h) $(\alpha_0, \eta_M, \alpha_1, \theta_M, L_0, L_1, R_{MH}) = (5', 50', 15', 20.6°, 10m, 2m, 9.47m)$.

**Figure 14** $AD_S$ for HFM at the end of an m=3 guide with $\eta_M=12'$, $\lambda=2.36$Å at $\theta_M=21.2°$

a) Flat monochromator
b) Focussed monochromator with $R_{MH}=9.51$ m
c) "Over"-Focussed monochromator with $R_{MH}=4.75$ m
d) "Over"-Focussed monochromator with $R_{MH}=3.17$ m showing reversed $AD_S$ slope.

**Figure 15** Schematic of novel primary spectrometer using open beamtubes, a horizontally curved mosaic monochromator, a variable width "virtual" source heavy slit with variable position between source and monochromator and a variable width heavy slit before the monochromator as discussed in section 8.